\documentclass[11pt]{article}

\usepackage[utf8]{inputenc}
\usepackage[T1]{fontenc}
\usepackage{lmodern}
\usepackage{microtype}

\usepackage{geometry}
\usepackage{setspace}
\usepackage{amsmath,amssymb,amsfonts,amsthm}
\usepackage{mathtools}
\usepackage{bm,bbm}

\usepackage{graphicx}
\usepackage{booktabs,longtable, float, array,tabularx, multirow}
\usepackage{booktabs}
\usepackage{siunitx}
\usepackage{placeins} 
\usepackage{float}
\usepackage{placeins} 
\usepackage{caption}
\usepackage{subcaption}
\usepackage{hyperref}
\usepackage{xcolor}
\usepackage{seqsplit}
\usepackage{pdflscape}
\usepackage{tikz}
\usetikzlibrary{arrows.meta, calc, positioning, shapes.geometric}
\usepackage{graphicx}
\usepackage{float}
\usepackage{caption}
\usepackage{subcaption}

\usepackage[dvipsnames]{xcolor}

\usepackage{enumitem}
\usepackage{verbatim}

\usepackage{algorithm}
\usepackage{algpseudocode}

\usepackage{tikz}
\usetikzlibrary{arrows.meta,calc,positioning}

\usepackage{comment}
\usepackage{marginnote}
\usepackage{todonotes}

\usepackage{appendix}
\usepackage{authblk}
\usepackage{ifthen}
\usepackage{xr}

\usepackage[most]{tcolorbox}
\tcbuselibrary{skins,breakable}

\newtcolorbox{stepbox}{%
  enhanced, breakable,
  colback=black!1, colframe=black, boxrule=0.6pt,
  arc=2pt, left=6pt, right=6pt, top=6pt, bottom=6pt,
  before skip=6pt, after skip=6pt,
}

\newcounter{workflowstep}

\usepackage[nameinlink,capitalise]{cleveref}
\usepackage{doi}
\usepackage[natbibapa]{apacite}

\newcommand{\argmin}{\operatorname*{arg\,min}}

\newcommand{\one}{\mathbf{1}}
\newcommand{\TE}{\ensuremath{\mathrm{TE}}}

\newcommand{\bp}{\ensuremath{\,\mathrm{bp}}}

\newcommand{\wold}{w^{\mathrm{old}}}
\newcommand{\wnew}{w^{\mathrm{new}}}
\newcommand{\wmap}{w_{\mathrm{MAP}}}

\newcommand{\epsinit}{\varepsilon_0}

\def\boxit#1{\smash{\fboxsep=0pt\llap{\rlap{\fbox{\strut\makebox[#1]{}}}~}}\ignorespaces}

\title{Low-Turnover Rebalancing for Sparse Index Tracking}

\author[]{Dimitrios Roxanas}

\affil[]{School of Mathematical and Physical Sciences,\\
The University of Sheffield, S3 7RH, United Kingdom.\\
Email: d.roxanas@sheffield.ac.uk}
\date{}

\begin{document}
\maketitle
\begin{abstract}

Sparse index tracking is often evaluated through rolling reconstruction:\ a sparse portfolio is fitted on an in-sample window, held over the next period, and rebuilt when the window rolls forward.\ This can achieve low realised tracking error, but it treats rebalancing primarily as repeated construction and can generate large turnover and frequent substitutions in the selected constituents.\ We propose a new workflow that separates sparse-tracker construction from sparse-tracker maintenance.\ A hybrid optimisation-plus-sampling framework provides the metrics operating at the decision level for both layers.\ The initial tracker is built from a calibrated shrinkage model and uncertainty--aware posterior support screening.\ Subsequent rebalance dates are handled in the self-financing change variable $\Delta w$.\ The default action is to preserve the existing tracker;\ local repairs are implemented only when realised tracking deterioration and posterior directional evidence jointly suggest intervention.\ In a 2020--2025 S\&P~500-style case study, we show that the proposed tracker occupies a distinct low-turnover operating region.\ Moreover, we demonstrate that the proposed $\Delta w$ maintenance layer can be attached to externally constructed trackers, where it gives consistent improvements over simply holding the initial tracker.
\end{abstract}



\tableofcontents

\section{Introduction}

Index-tracking funds, a popular passive asset management instrument, seek to reproduce the performance of a market index using a portfolio of its constituents.\ In practice, investors rarely hold all the names in an index:\ partial replication reduces administrative overhead, avoids tiny or illiquid positions, and limits the rebalancing burden.\ Mathematically, this gives a high-dimensional optimisation problem in which tracking error (TE) must be balanced against sparsity, budget, implementability, and turnover considerations.\
Throughout, we work with target portfolio weights and report turnover as the main implementation-cost diagnostic;\ the precise value-weight accounting convention is stated in \Cref{sec:methodological-overview}.

A common empirical practice is rolling-window reconstruction:\ fit a sparse portfolio on a historical window, hold it over the next out-of-sample period, then refit when the window rolls forward.\ This convention is natural because index composition, relative capitalisations, and market dynamics evolve.\ It also means, however, that a method which is excellent as a \emph{construction engine} may behave very differently as a \emph{maintenance policy}.\ Re-solving for a low-TE sparse portfolio at each date can repeatedly alter the selected support, producing short holding spells, round-trip trades, and high turnover.\ The complementary question studied here is therefore operational:\ given an incumbent sparse tracker, when is there enough evidence to spend turnover on a local repair rather than leave the tracker unchanged or reconstruct it from scratch?\ Our focus on turnover is because of its use as an implementation-cost proxy:\ proportional transaction and slippage costs are commonly approximated as being linear in it.\ 

We propose a workflow that combines regularised optimisation, calibrated generalised-Bayes updating, empirical-Bayes shrinkage calibration, proximal MCMC uncertainty summaries, and posterior-guided portfolio decisions.\ The methodological object is not a single estimator but a maintenance pipeline:\ construct a sparse tracker, quantify uncertainty in its effective support, monitor realised deterioration, and apply small self-financing repairs only when posterior evidence supports them.\ In particular:

\begin{enumerate}[label=(A\arabic*),leftmargin=*]
\item rather than relying on a single exact combinatorial optimisation stage, we formulate sparse index tracking and rebalancing as a hybrid computational pipeline that combines shrinkage, approximate posterior exploration, and posterior-guided decision rules;

\item we introduce a calibrated shrinkage--selection mechanism in which an effective tracking-loss scale is selected, a global weighted-Laplace sparsity rate is estimated conditionally using the stochastic approximation proximal gradient (SAPG) scheme, and the resulting maximum a posteriori estimator (MAP)/posterior pair is used as the input to a separate support-decision layer;

\item we decouple shrinkage from final selection: a continuous sparse model is learned first, then posterior magnitude thresholds and activation probabilities are used to determine effective holdings and justified rebalancing moves;

\item we use the resulting posterior information to support conservative rebalancing decisions and, when realised tracking error deteriorates relative to a target band, perform a self-financing posterior-informed local repair rather than reconstructing the portfolio;

\item decompose the empirical comparison into static holding, low-turnover maintenance, externally supplied starting trackers, and rolling reconstruction, so that tracking error is interpreted jointly with turnover and support persistence.
\end{enumerate}

We illustrate the approach on a case study tracking the S\&P~500 index using a universe of several hundred constituents over multiple fitting and holding periods.\ The case study is used as an integral part of the methodological demonstration:\ it shows (i) how the sparse tracker is constructed, (ii) how posterior information shapes the effective support, (iii) how posterior uncertainty can gate rebalancing decisions, and (iv) how the same machinery behaves under alternative evaluation modes.\ Existing sparse reconstruction baselines are allowed to rebuild the portfolio on each date;\ the proposed method instead inspects the incumbent tracker and trades only through approved local repairs.\ Additional experiments ask what happens if the tracker is never maintained, if the $\Delta w$ layer is plugged onto external trackers, and if the uncertainty quantification (UQ) construction procedure is forced into the same rolling-reconstruction regime as external trackers.\\

Additional diagnostics, sensitivity experiments, and computational details are reported in the companion Supplementary Material.\ Replication code and logs of several experiments are available at
\href{https://github.com/droxanas/Low-Turnover-Rebalancing-for-Sparse-Index-Tracking/}{https://github.com/droxanas/Low-Turnover-Rebalancing-for-Sparse-Index-Tracking/}.

\section{Related Work}

\paragraph{Sparse construction and rolling reconstruction}

The classical sparse index-tracking problem minimises tracking error under cardinality, budget, and implementability constraints.\ Mixed-integer programming (MIP) formulations and evolutionary heuristics have long been used for this task \cite{Beasley2003, CanakgozBeasley2009}.\ More recent continuous optimisation methods, including the Majorisation--Minimisation (MM) framework of \citet{Benidis2018a,Benidis2018b}, provide strong high-dimensional sparse-reconstruction approaches.\ Penalised-regression approaches such as non-negative LASSO (NNL) and non-negative elastic net (NNEN) are also natural construction baselines because they combine variable selection and non-negativity in a regression representation of index tracking \cite{wu2014nonnegativelasso,wu2014nonnegativeEN}.\ These methods are highly effective as construction or reconstruction engines:\ given a fitting window and a target sparsity level, they produce a new sparse portfolio designed to track the benchmark closely.\ Their rolling empirical use, however, often amounts to repeatedly replacing the sparse portfolio.\ From the point of view of the present paper, this is not a flaw in those methods;\ it is a different decision mode.\ We therefore report rolling reconstruction baselines as the high-accuracy/high-turnover reference regime rather than as the only relevant benchmark.\ Comprehensive reviews document the breadth of optimisation, statistical, and data-driven formulations \cite{Silva2024,dhingra2026index}.\

\paragraph{Rebalancing, transaction costs, and dynamic tracking}

Several operations research papers treat rebalancing as a problem in its own right.\ \citet{Gaivoronski2005} explicitly distinguish static and dynamic benchmark tracking and propose a threshold rule that rebalances only when a newly computed candidate portfolio improves sufficiently over the current one after accounting for transaction costs.\ \citet{CanakgozBeasley2009} formulate index tracking and enhanced indexation as mixed-integer linear programs with cardinality constraints, transaction costs, and transaction-cost budgets, and evaluate both single-split and systematic revision protocols.\ \citet{Strub2018} develop a mixed-integer formulation for construction and rebalancing with cash deposits and withdrawals, explicitly managing the trade-off between tracking accuracy and transaction costs.\ \citet{BarroCanestrelli2009} study multistage benchmark replication with transaction costs and liquidity in a stochastic-programming framework.\  \citet{HuangZhangZhao2021} motivate multi-period index tracking by observing that independent single-period solutions can lead to frequent rebalancing and high transaction costs.\ The work \citet{Chiam2013} formulates dynamic index tracking as a multi-objective evolutionary optimisation problem that trades off tracking performance and transaction costs under practical constraints, and compares no rebalancing, periodic rebalancing, and event-triggered tracking-error-limit rebalancing.\ Our method differs in the mechanism:\ rather than re-optimising a full tracker via a multi-objective evolutionary search whenever a trigger fires, we treat the incumbent sparse tracker as the state and use posterior directional evidence to approve small local $\Delta w$ repairs.

\paragraph{Turnover sparsity and sparse updates}

A particularly close recent reference is \citet{YamagataOno2024}, who explicitly distinguish constraints on portfolio sparsity from constraints on turnover sparsity.\ Their formulation allows an $\ell_0$ constraint either on the portfolio itself or on the update from the previous portfolio, thereby recognising that sparse holdings and sparse trades are different objects.\ This distinction is central to the present work.\ The difference is that our sparse-update mechanism is not only an optimisation constraint:\ candidate changes are generated by a $\Delta w$ MAP, filtered by posterior directional probabilities, and implemented only under realised tracking deterioration and a hard turnover budget.

\paragraph{Shrinkage, generalised Bayes, and posterior computation}

The regression view of index tracking connects the problem to sparse estimation.\ LASSO, non-negative LASSO, elastic-net, and related shrinkage methods provide natural sparse-regression baselines \cite{tibshirani1996regression, Zou2005elastic,wu2014nonnegativelasso,wu2014nonnegativeEN}.\ A Bayesian reading of such penalties treats them as priors, for example the Laplace prior underlying the Bayesian LASSO \cite{ParkCasella2008}.\ In sparse index tracking, however, the Gaussian regression likelihood should not be interpreted literally:\ a sparse tracker is an approximation to a benchmark, not a generative model for the benchmark return.\ We therefore use the \emph{generalised-Bayes} view, in which a posterior-like update is built from an exponentiated loss rather than a fully specified sampling model \cite{bissiri2016general}.\ Conditional on a calibrated loss scale, the global weighted-Laplace sparsity rate is estimated by SAPG, following empirical-Bayes regularisation-parameter methods for high-dimensional inverse problems \cite{Vidal2020}.

Posterior computation is used here not to replace optimisation but to turn a sparse optimisation model into a decision-support mechanism.\ The MAP is computed with proximal optimisation, while posterior summaries are obtained with proximal Langevin-type MCMC methods (MYULA and MALA) based on Moreau--Yosida smoothing of nonsmooth penalties \cite{pereyra2016proximal, DurmusMoulinesPereyra2018, Zygalakis2019}.\ These summaries are converted into transparent portfolio rules: support selection, implementability filtering, and directional gates for local repairs.

\paragraph{Evaluation-protocol implication}

The literature does not impose a single universal empirical protocol.\ Penalised-regression papers often compare internal variants such as NNL-plus-OLS versus NNL alone, or NNEN versus NNL.\ MIP and evolutionary rebalancing papers typically compare modelling variants, transaction-cost settings, selection rules, rebalancing frequencies, or event-triggered policies within their own framework.\ Our evaluation follows the same principle:\ the comparison is designed to illuminate the modelling claim.\ We therefore distinguish static holding, low-turnover maintenance, rolling reconstruction, and plug-in maintenance from external starting trackers.\

\paragraph{Methodological comparison and positioning}

Full reconstruction methods solve for a new target portfolio $w^{\mathrm{new}}$ on each window, sometimes controlling turnover by penalising distance from current holdings.\ Our sequential layer instead works directly with the self-financing trade
\[
    \Delta w = w^{\mathrm{new}} - w^{\mathrm{old}},\qquad \mathbf{1}^{\top}\Delta w=0,
\]
and places the sparsity and uncertainty analysis on this change vector.\ The default action is inaction.\ A trade is implemented only when the MAP direction, posterior directional probability, realised-TE monitoring rule, long-only feasibility, and turnover budget jointly support a bounded local repair.\ This distinguishes the framework from both periodic reconstruction and deterministic turnover-penalised reconstruction:\ the proposed method contributes an uncertainty-gated maintenance layer for sparse passive tracking.

\section{Methodological overview}\label{sec:methodological-overview}

The proposed workflow treats sparse index tracking as a computational decision problem rather than as the recovery of a true sparse index composition.\ The benchmark index is not itself sparse:\ a capitalisation-weighted benchmark assigns nonzero weight to many constituents, and a sparse tracker is necessarily an approximation.\ For this reason, the constrained regression representation used below should be read as a tracking-loss device, not as a literal generative model for market-index returns.\ This distinction is important for this paper.\ It allows the variance-like scale in the quadratic loss to be calibrated for portfolio-construction quality, while retaining the useful Bayesian language of shrinkage, posterior exploration, and uncertainty-informed decisions.

The accounting convention is as follows.\ A reported portfolio \(w\) is a vector of target value weights, not share counts: \(w_j\) denotes the fraction of current portfolio value allocated to asset \(j\).\ The implemented portfolios in
the case study are long-only and fully invested, i.e.,
\[
    w_j \ge 0, \qquad \sum_{j=1}^p w_j = 1.
\]
If \(w^{\mathrm{old}}\) is updated to \(w^{\mathrm{new}}\) by a trade vector
\(\Delta w = w^{\mathrm{new}}-w^{\mathrm{old}}\), then a self-financing update satisfies
\[
    \sum_{j=1}^p \Delta w_j = 0.
\]
Thus, buys must be funded by sells or by controlled reductions in existing holdings in value-weight terms.\ The construction posterior below enforces the budget condition softly for computational reasons;\ the final reported portfolios and implemented trades enforce the accounting constraints explicitly.

On a fitting window, let \(y\in\mathbb{R}^T\) denote the benchmark return vector and let \(R\in\mathbb{R}^{T\times p}\) denote the asset-return matrix.\ After centring the data, the construction loss is
\begin{equation}
  \ell(w;y,R)
  =
  \frac12\|y_c-R_cw\|_2^2.
  \label{eq:method-loss}
\end{equation}
We use a generalised-Bayes posterior of the form
\begin{equation}
  \pi_{\omega,\theta}(w\mid y,R)
  \propto
  \exp\left\{
     -\omega\,\ell(w;y,R)
     -\theta \sum_{j=1}^p \alpha_j |w_j|
     -\Lambda(\one^\top w-1)^2
  \right\}.
  \label{eq:method-generalised-posterior} 
\end{equation}
Here \(\omega\) is a learning rate, or inverse temperature, for the tracking loss;\ \(\theta\) is the global weighted-Laplace sparsity rate;\ \(\alpha_j\) are column-scale weights; and \(\Lambda\) controls a soft penalty for deviations from the fully invested budget condition \(\sum_j w_j=1\).\ The soft budget term is used inside the construction-stage posterior exploration, not as the final accounting rule for the investable portfolio.\ It allows the shrinkage posterior to be explored in ambient coordinates while keeping posterior mass close to the fully invested hyperplane.\ After support selection, the portfolio is refitted under exact long-only and fully invested constraints.\ We emphasise that \(\omega\) is not estimated as a physical noise precision.\ Instead, it indexes a family of models that trade tracking fit, shrinkage, and computational stability.\ Conditional on each candidate value of \(\omega\), the sparsity rate \(\theta\) is calibrated by the SAPG empirical-Bayes scheme (details in the Appendix and the Supplementary Material), and the corresponding MAP is computed by proximal optimisation (FISTA, \cite{beck2009fista}).\ The resulting pair
\[
 \, \left(w_{\mathrm{MAP}}(\omega),\,\pi_{\omega,\theta(\omega)}(w\mid y,R)\,\right)
\]
therefore defines one candidate sparse-tracking model.

The outer calibration step chooses among these candidates by a low-dimensional portfolio score.\ The score selects an effective loss scale that produces an acceptable TE--sparsity compromise.\ 
This is the first decoupling in the workflow:\ shrinkage is used to create a posterior landscape with ``strong'' and ``weak'' coordinates, but it is not expected to deliver the final implementable support by itself.\ This is natural for a continuous weighted-Laplace specification.\ Although the underlying nonsmooth weighted-\(\ell_1\) optimisation problem can in principle produce exact zeroes, the posterior itself does not contain discrete inclusion indicators, and small or borderline coordinates may be sensitive to scale, correlation, smoothing and numerical thresholding.\ 
The transition from a shrinkage-based posterior/MAP description to a tradable support is therefore treated as a separate decision problem.

The second layer converts posterior information into support decisions.\ Let \(\wmap\) be the MAP selected by the loss-scale/SAPG calibration step, and let
\[
  \{w^{(m)}\}_{m=1}^M\]
be post-burn draws from the corresponding posterior.\ For an implementability threshold \(\epsinit>0\), define the practical-positive posterior probability
\begin{equation}
  \widehat p_j^+(\epsinit)
  =
  \frac1M\sum_{m=1}^M
  \mathbbm{1}\{w_j^{(m)} > \epsinit\}
  \approx
  \mathbb{P}(w_j>\epsinit\mid y,R).
  \label{eq:method-practical-positive}
\end{equation}
The initial support is then selected by
\begin{equation}
  S_{\mathrm{init}}(\epsinit,\pi_+^*)
  =
  \left\{
    j:
    w_{\mathrm{MAP},j}>\epsinit,
    \quad
    \widehat p_j^+(\epsinit)\geq \pi_+^*
  \right\}.
  \label{eq:method-support-rule}
\end{equation}
Thus, the MAP supplies the magnitude and sign information, while the posterior sample supplies evidence that the coordinate is practically positive rather than merely numerically nonzero.\ The selected names are then refitted under the final implementability constraints, including exact budget and long-only constraints.\ In this way, the posterior rule selects the support, and the final constrained refit determines the portfolio weights.

The same decision principle is used in the sequential maintenance layer, but the unknown is now the trade rather than the full portfolio.\ At a rebalance date \(h\), the current portfolio \(\wold_h\) is treated as the state variable and candidate changes are constrained to be self-financing,
\begin{equation}
  \wnew_h = \wold_h + \Delta w_h,
  \qquad
  \one^\top \Delta w_h = 0.
  \label{eq:method-self-financing-trade}
\end{equation}
A sparse generalised posterior is then built on \(\Delta w_h\), using the latest fitting window and a weighted penalty on the change vector.\ A coordinate of \(\Delta w_h\) is considered only when the \(\Delta w\)-MAP has practically non-negligible magnitude and the posterior sample supports the MAP direction.\ For example, for a proposed positive move, the directional probability is
\begin{equation}
  \widehat p_{\mathrm{dir},h,j}(\varepsilon)
  =
  \frac1M\sum_{m=1}^M
  \mathbbm{1}\{\Delta w_{h,j}^{(m)} > \varepsilon\},
  \label{eq:method-directional-prob-positive}
\end{equation}
with the analogous left-tail probability for a proposed negative move.\ Recovery decisions are then based on MAP magnitude, posterior directional evidence, and realised tracking-error deterioration.\ Approved moves are implemented as bounded local repairs, with buys and sells matched locally, whenever possible, and without globally renormalising the entire portfolio.

This gives the workflow its main methodological character.\ The construction stage asks which sparse tracker should be used initially.\ The support-selection stage asks which posterior-supported coordinates are practically worth holding.\ The maintenance stage asks whether the current tracker should be changed at all, and if so, which small self-financing repair is supported by posterior directional evidence.\ \Cref{sec:workflow} records the concrete effective-variance grids, scoring rules, support thresholds, and recovery-mode controls used in the empirical study.

\section{Proposed workflow}\label{sec:workflow}

This section turns the methodological overview into the concrete workflow used in the empirical study.\ The workflow has two linked but distinct parts.\ The construction layer selects an initial sparse tracker from a family of calibrated generalised-Bayes models.\ The maintenance layer treats the current tracker as a state variable and asks whether a small self-financing trade is justified.\

\begin{figure}[!ht]
\centering
\includegraphics[height=7.8cm,keepaspectratio]{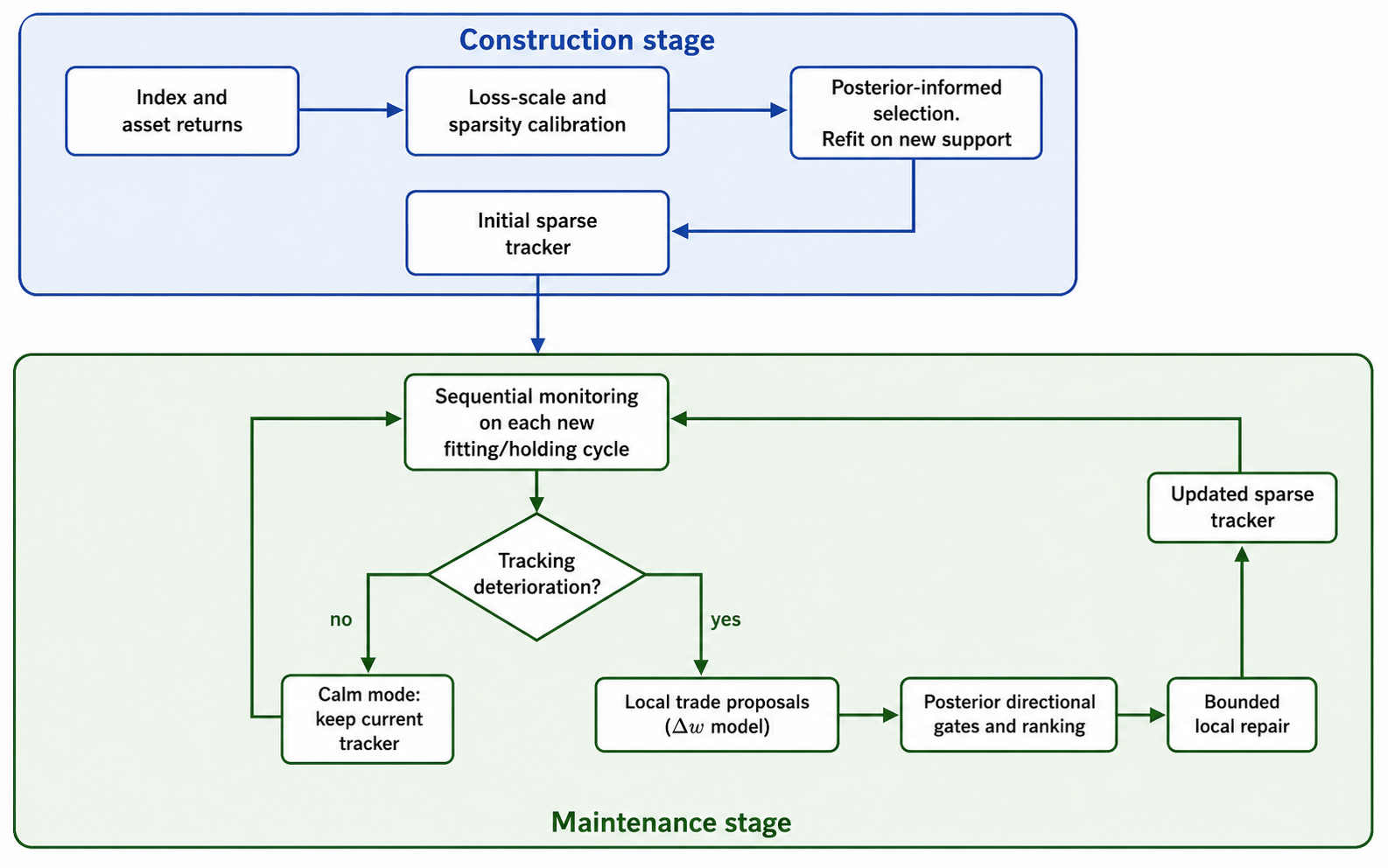}
\caption{Construction and low-turnover maintenance workflow.\ The tracker is built once, then either kept unchanged or locally repaired through posterior-gated self-financing \(\Delta w\) updates.}
\end{figure}

\subsection{Operational summary}\label{subsec:workflow-summary}

The empirical implementation can be summarised as the following sequence.

\begin{enumerate}[label=\textbf{Step \arabic*.}, leftmargin=2.2cm]
  \item \textbf{Prepare the data on the fitting window.} For a fitting window of length $T$, form centred benchmark and asset returns \((y_c,R_c)\), and compute mean-normalised column scales \(\alpha_j\propto\|R_{c,\cdot j}\|_2/\sqrt{T}\).

  \item \textbf{Scan construction loss scales.} For each multiplier \(c\) in a finite grid, form an effective variance \(\sigma^2_{\mathrm{eff}}(c)\), estimate the weighted-Laplace sparsity rate by SAPG, conditional on that scale, and compute the corresponding sparse MAP.\ Select the construction row using a low-dimensional TE--sparsity score.

  \item \textbf{Select and refit the initial support.}\ Explore the selected posterior by MCMC, retain names that are practically positive with sufficient posterior probability, and refit the chosen support under simplex constraints.

  \item \textbf{Apply a one-off implementability floor.}\ Before the first holding window (only), remove sub-floor ``dust'' holdings and redistribute their mass over the remaining support.\ This is an implementation adjustment, not a repeated rebalancing rule.

  \item \textbf{Monitor the held tracker.}\ At each subsequent rebalance date, compute realised tracking diagnostics for the currently held portfolio.\ The default action is to keep the existing portfolio unchanged.

  \item \textbf{Solve a local \(\Delta w\) problem.}  At a rebalance date, solve a self-financing sparse change problem over a \(\Delta w\) loss-scale grid, akin to {\bf Step 2}.\ The grid selects a candidate MAP trade and posterior sample, not a new full portfolio.

  \item \textbf{Gate and implement trades.}  Candidate moves are filtered by posterior directional probabilities.\ ``Calm'' mode uses stricter gates and usually does not trade.\ ``Recovery'' mode is triggered by realised TE deterioration and implements only bounded, locally funded repairs.
\end{enumerate}

The posterior sampling summaries used below are computed with a proximal Langevin/MALA implementation.\ Since the nonsmooth weighted-$\ell_1$ term is handled through a Moreau--Yosida regularisation, Appendix \ref{app:computational-summary} records the precise \emph{smoothed} target, proposal, and Metropolis correction for the MALA implementation.\ In the main workflow, we suppress this technical distinction and refer simply to the MAP estimate and posterior sample. 

The constants below are the fixed settings for the full-horizon case study;\ they are reported for transparency, and they should not be seen as ``optimal''.\ The main methodological point is the \emph{structure} of the decision rule, i.e., calibrated construction followed by uncertainty-gated maintenance.

\subsection{Construction-stage loss-scale grid}\label{subsec:workflow-construction-grid}

On the initial fitting window, the baseline scale is the centred (estimated, e.g., by MAD) OLS residual variance \(\widehat\sigma^2_{\mathrm{OLS},0}\).\ The construction layer scans
\begin{equation}
  \mathcal C
  =\{1,10,25,50,75,100,125,150,200\},
  \qquad
  \sigma^2_{\mathrm{eff},0}(c)=c\,\widehat\sigma^2_{\mathrm{OLS},0}.
  \label{eq:workflow-construction-c-grid}
\end{equation}
For each \(c\in\mathcal C\), SAPG estimates a conditional empirical-Bayes rate \(\theta_{\mathrm{EB}}(c)\).\ The full-weight construction MAP is
\begin{equation}
  w_{\mathrm{MAP}}(c)
  \in
  \argmin_{w\in\mathbb R^p}
  \left\{
  \frac{1}{2\sigma^2_{\mathrm{eff},0}(c)}\|y_c-R_cw\|_2^2
  +\Lambda(\one^\top w-1)^2
  +\theta_{\mathrm{EB}}(c)\sum_{j=1}^p\alpha_j|w_j|
  \right\}.
  \label{eq:workflow-construction-map}
\end{equation}
In the implementation, \(w_{\mathrm{MAP}}(c)\) denotes the minimiser of the Moreau--Yosida-smoothed counterpart of \eqref{eq:workflow-construction-map}.\ We display the underlying nonsmooth objective to identify the tracking, budget, and weighted-\(\ell_1\) terms being smoothed.

The above is an effective loss-scale grid:\ changing \(c\) changes the relative strength of the tracking loss, the soft budget penalty, and the fitted sparse-prior rate.\
The construction score combines a loose TE feasibility gate with an effective-sparsity preference.\ Define
\begin{equation}
  n_{\mathrm{eff}}(c)
  =\#\{j:|w_{\mathrm{MAP},j}(c)|\geq 10^{-4}\},
  \label{eq:workflow-construction-neff}
\end{equation}
and
\begin{equation}
  \phi^{\mathrm{FIT}}_{\mathrm{TE}}(c)
  =
  \mathbbm 1\!\left\{
    \gamma_{\mathrm{lo}}\TE_{\mathrm{ref},0}
    \leq
    \TE_{\mathrm{FIT}}\{w_{\mathrm{MAP}}(c)\}
    \leq
    \gamma_{\mathrm{hi}}\TE_{\mathrm{ref},0}
  \right\},
  \;\;
  \gamma_{\mathrm{lo}}=0,\quad \gamma_{\mathrm{hi}}=5,
  \label{eq:workflow-construction-te-gate}
\end{equation}
where \(\mathrm{TE}_{\mathrm{ref},0}\) is the OLS tracking-error reference on the initial fitting window. (with a small numerical lower floor \(10^{-5}\)) 

The target effective cardinality before posterior gating is chosen to be \(n_{\mathrm{FIT}}^\star=150\), with ``bandwidth'' \(s_{\mathrm{nnz}}=25\), giving
\begin{equation}
  w^{\mathrm{FIT}}_{\mathrm{nnz}}(c)
  =
  \exp\left\{-\frac12
  \left(
  \frac{n_{\mathrm{eff}}(c)-n_{\mathrm{FIT}}^\star}{s_{\mathrm{nnz}}}
  \right)^2
  \right\},
  \qquad
  d_{\mathrm{FIT}}(c)=\phi^{\mathrm{FIT}}_{\mathrm{TE}}(c)w^{\mathrm{FIT}}_{\mathrm{nnz}}(c).
  \label{eq:workflow-construction-score}
\end{equation}
The branch used in the empirical study is the selected \(c^\star=25\) row.\ 

\subsection{Posterior support selection, refit, and implementability floor}\label{subsec:workflow-support-refit-floor}

Let \(w_{\mathrm{MAP}}=w_{\mathrm{MAP}}(c^\star)\), and let \(\{w^{(m)}\}_{m=1}^M\) be post-burn samples from the corresponding posterior.\ For \(\varepsilon_0=10^{-3}\), we define
\begin{equation}
  \widehat p_j^+(\varepsilon_0)
  =
  \frac1M\sum_{m=1}^M
  \mathbbm 1\{w_j^{(m)}>\varepsilon_0\}.
  \label{eq:workflow-practical-positive}
\end{equation}
The initial active set is
\begin{equation}
  S_{\mathrm{init}}
  =
  \left\{
  j:
  w_{\mathrm{MAP},j}>10^{-3},\quad
  \widehat p_j^+(10^{-3})\geq 0.60
  \right\}.
  \label{eq:workflow-initial-support}
\end{equation}
This rule deliberately separates shrinkage from final selection.\ The continuous weighted-Laplace model supplies a MAP and posterior sample, but the implementable support is chosen only after asking whether each coordinate is practically positive with sufficient posterior evidence.

The selected names are then refitted under simplex constraints,
\begin{equation}
  \widehat w_{S_{\mathrm{init}}}
  \in
  \argmin_{v\in\mathbb R^{|S_{\mathrm{init}}|}}
  \|y_c-R_{c,S_{\mathrm{init}}}v\|_2^2,
  \qquad
  v\geq0,
  \quad
  \one^\top v=1,
  \label{eq:workflow-support-refit}
\end{equation}
with all non-selected coordinates set to zero.\  Finally, the case-study branch applies a one-off floor \(\ell_{\mathrm{floor}}=0.005\).\ If
\begin{equation}
  F=\{j:\widehat w_j\geq \ell_{\mathrm{floor}}\},
\end{equation}
then the implemented initial portfolio is
\begin{equation}
  w^{(0)}_j
  =
  \begin{cases}
  \widehat w_j\Big/\sum_{\ell\in F}\widehat w_\ell, & j\in F,\\[4pt]
  0, & j\notin F.
  \end{cases}
  \label{eq:workflow-floor-map}
\end{equation}
The floor is not applied repeatedly.\ It prevents the empirical case study from relying on dust-level holdings while preserving the posterior-selected support logic.\ The specific floor level is of the same order as the one used in \citet{Benidis2018b} for the lower bound of their holding constraint.

\subsection{Sequential \(\Delta w\) loss-scale grid}\label{subsec:workflow-dw-grid}

At a later rebalance date \(h\), the current portfolio \(w_h^{\mathrm{old}}\) is treated as fixed.\ Candidate updates are self-financing changes,
\begin{equation}
  w_h^{\mathrm{new}}=w_h^{\mathrm{old}}+\Delta w_h,
  \qquad
  \one^\top\Delta w_h=0.
  \label{eq:workflow-dw-self-financing}
\end{equation}
The equality \(1^\top \Delta w_h=0\) is the value-weight expression of self-financing accounting.\ Both \(w_h^{\mathrm{old}}\) and \(w_h^{\mathrm{new}}\) are target portfolio weights, expressed as fractions of current portfolio value.\ Thus, an approved increase in some weights must be matched by equal total reductions in other weights.\ Our case study abstracts from external cash flows and from deducting transaction costs at rebalance dates, so the maintenance layer operates by locally reallocating existing portfolio weights and reports turnover separately as an implementation-cost diagnostic.\

The final executed change must also preserve long-only feasibility.\ Thus, the admissible implementation set is
\begin{equation}
  \mathcal A_h
  =
  \left\{
  u\in\mathbb R^p:
  \one^\top u=0,
  \quad
  w_h^{\mathrm{old}}+u\geq0
  \right\}.
  \label{eq:workflow-admissible-change-set}
\end{equation}
Negative components of a change vector are therefore allowed, but only as sales from existing positive holdings; they cannot create short positions.\ The \(\Delta w\)-MAP below is used to propose directions and priorities, while the bounded implementation rule in \Cref{subsec:workflow-local-implementation} enforces feasibility for the realised trade.

The rebalancing layer uses an effective variance scale directly, rather than first estimating a residual variance for the change problem:\
\begin{equation}
  \sigma^2_{\Delta,\mathrm{base}}=10^{-4},
  \qquad
  \sigma^2_{\Delta,h}(c)=c\,10^{-4},
  \qquad c\in\mathcal C.
  \label{eq:workflow-dw-c-grid}
\end{equation}
For each \(c\), SAPG estimates \(\kappa_{\mathrm{EB},h}(c)\) conditional on the sum-zero change space.\ The \(\Delta w\)-MAP row is based on
\begin{equation}
  \Delta w_{\mathrm{MAP},h}(c)
  \approx
  \argmin_{\one^\top u=0}
  \left\{
  \frac{1}{2\sigma^2_{\Delta,h}(c)}
  \|y_{c,h}-R_{c,h}(w_h^{\mathrm{old}}+u)\|_2^2
  +\kappa_{\mathrm{EB},h}(c)
  \sum_{j=1}^p \alpha^{(\Delta)}_j|u_j|
  \right\}.
  \label{eq:workflow-dw-map}
\end{equation}
As in the construction stage, the implemented \(\Delta w\)-MAP is obtained from the Moreau--Yosida-smoothed version of this objective.\ The rebalancing scales are
\begin{equation}
  \alpha^{(\Delta)}_j
  =
  \frac{\left(\alpha^{\mathrm{base}}_j\right)^2}
       {p^{-1}\sum_{\ell=1}^p\left(\alpha^{\mathrm{base}}_\ell\right)^2},
  \qquad
  \alpha^{\mathrm{base}}_j\propto\frac{\|R_{c,h,\cdot j}\|_2}{\sqrt T}.
  \label{eq:workflow-dw-alpha}
\end{equation}
The squared scale in \eqref{eq:workflow-dw-alpha} makes the change penalty more conservative (compared to the design stage) for assets whose fitting-window returns carry a larger column scale, essentially asking for stronger posterior evidence to justify a change.

Also, in contrast to the original design stage, the selected row is chosen by an improvement--movement score rather than by a target number of moved names.\ Let
\begin{equation}
  \TE^{\mathrm{new}}_h(c)
  =
  \TE_{\mathrm{FIT},h}\{w_h^{\mathrm{old}}+\Delta w_{\mathrm{MAP},h}(c)\},
  \qquad
  I_h(c)=
  \frac{\TE^{\mathrm{old}}_h-\TE^{\mathrm{new}}_h(c)}{\TE^{\mathrm{old}}_h}.
  \label{eq:workflow-dw-improvement}
\end{equation}
The TE component is
\begin{equation}
  w_{\TE,h}(c)
  =
  \min\left\{1,\frac{[I_h(c)]_+}{0.05}\right\},
  \label{eq:workflow-dw-te-score}
\end{equation}
so a row receives full TE credit after a 5\% fitting-window TE improvement.\ The movement component is
\begin{equation}
  L^{\Delta}_{1,h}(c)=\|\Delta w_{\mathrm{MAP},h}(c)\|_1,
  \qquad
  w_{\ell_1,h}(c)
  =
  \exp\left\{-\frac12
  \left(
  \frac{[L^{\Delta}_{1,h}(c)-0.03]_+}{0.02}
  \right)^2
  \right\},
  \label{eq:workflow-dw-l1-score}
\end{equation}
and the grid score is
\begin{equation}
  d_{\Delta,h}(c)=w_{\TE,h}(c)w_{\ell_1,h}(c).
  \label{eq:workflow-dw-score}
\end{equation}
Thus, \(\|\Delta w\|_1\leq0.03\), approximately 1.5\% one-way turnover, is not penalised by the score, while larger proposed movements are softly discouraged.\ 

\subsection{Posterior directional gates}\label{subsec:workflow-directional-gates}

After the winning \(\Delta w\) grid row has been selected, samples \(\{\Delta w_h^{(m)}\}_{m=1}^M\) from the MALA run are used to assess marginal directional evidence.\ For \(\varepsilon\geq0\), define
\begin{equation}
  \widehat p_{\mathrm{dir},h,j}(\varepsilon)
  =
  \begin{cases}
  M^{-1}\sum_{m=1}^M\mathbbm 1\{\Delta w_{h,j}^{(m)}>\varepsilon\},
  & \Delta w_{\mathrm{MAP},h,j}>0,\\[4pt]
  M^{-1}\sum_{m=1}^M\mathbbm 1\{\Delta w_{h,j}^{(m)}<-\varepsilon\},
  & \Delta w_{\mathrm{MAP},h,j}<0.
  \end{cases}
  \label{eq:workflow-directional-prob}
\end{equation}
The sign of the $j$-th MAP coordinate defines the proposed trade direction;\ the posterior sample measures how consistently that direction appears under the posterior.

In ``calm'' mode, candidate moves must pass a practical magnitude and posterior direction test,
\begin{equation}
  S_{\mathrm{calm},h}
  =
  \left\{
  j:
  |\Delta w_{\mathrm{MAP},h,j}|\geq10^{-4},\quad
  \widehat p_{\mathrm{dir},h,j}(10^{-4})\geq0.60
  \right\}.
  \label{eq:workflow-calm-set}
\end{equation}

For the case study we implement trades according to the following (example set) of rules:\ if both buys and sells are approved, the implementation is restricted to those names and matched locally.\ If only sells are approved, the trade is not executed because there is no posterior-approved destination for the released capital.\ If only buys are approved, a small set of weak existing holdings may be used as funders, subject to the same no-global-renormalisation principle used in recovery mode, that we discuss next.

\subsection{Recovery trigger and gating}\label{subsec:workflow-recovery}

Let \(\widehat\TE^{\mathrm{hold}}_{h-1}\) and \(\widehat\TE^{\mathrm{hold}}_{h-2}\) denote realised tracking errors, in basis points (bp)\footnote{\(1\,\mathrm{bp}=10^{-4}\).}, over the two most recent holding windows.\ A recovery inspection is triggered when either the previous realised tracking error exceeds a target band, or when the tracking error jumps sharply from one holding window to the next:
\begin{equation}
  \widehat\TE^{\mathrm{hold}}_{h-1}>17
  \qquad\text{or}\qquad
  \widehat\TE^{\mathrm{hold}}_{h-1}
  -
  \widehat\TE^{\mathrm{hold}}_{h-2}
  \geq7.
  \label{eq:workflow-recovery-trigger}
\end{equation}
Equivalently, the rule uses a 15 bp target level, a 2 bp tolerance gap, and a 7 bp shock-jump trigger.

For ease of reference, we define the set of active recovery anchors
\begin{equation}
  \mathcal R_h
  =
  \left\{15:\, \text{if}\;\widehat\TE^{\mathrm{hold}}_{h-1}>17\right\}
  \cup
  \left\{
  \widehat\TE^{\mathrm{hold}}_{h-2}:\, \text{if}\;
  \widehat\TE^{\mathrm{hold}}_{h-1}
  -
  \widehat\TE^{\mathrm{hold}}_{h-2}
  \geq7
  \right\}.
  \label{eq:workflow-recovery-anchor-set}
\end{equation}
Recovery mode is considered only when \(\mathcal R_h\neq\emptyset\).\ When this happens, the reference anchor is
\begin{equation}
  a_h=\min\mathcal R_h.
  \label{eq:workflow-recovery-anchor}
\end{equation}
Thus, \(a_h=15\) when the absolute target-band rule fires, \(a_h\) equals the previous holding-window TE when only the jump rule fires, and the smaller of the two reference levels is used when both rules fire.\ 

``Recovery'' mode changes the post-MALA decision rule, not the \(\Delta w\) grid.\ Recovery mode is entered only after realised tracking deterioration has already been diagnosed;\ at that point, the relevant decision is no longer whether to suppress ordinary low-confidence micro-trades, but whether there is enough directional evidence to spend a small, pre-specified repair budget.\ Thus, the thresholds below should be read as recovery-mode activation rules rather than as a claim of overwhelming marginal posterior certainty.\ The remaining safeguards---the MAP magnitude gate, ranking by posterior-weighted MAP size, truncation of the approved list, local funding, the long-only feasibility constraint, and the one-way turnover cap---control the size and scope of the implemented intervention.
\begin{equation}
  p^{\mathrm{req}}_{h,j}
  =
  \begin{cases}
  0.525, & w^{\mathrm{old}}_{h,j}>0\ \text{and}\ \Delta w_{\mathrm{MAP},h,j}>0,\\
  0.525, & w^{\mathrm{old}}_{h,j}>0\ \text{and}\ \Delta w_{\mathrm{MAP},h,j}<0,\\
  0.575, & w^{\mathrm{old}}_{h,j}=0\ \text{and}\ \Delta w_{\mathrm{MAP},h,j}>0,\\
  +\infty, & w^{\mathrm{old}}_{h,j}=0\ \text{and}\ \Delta w_{\mathrm{MAP},h,j}<0.
  \end{cases}
  \label{eq:workflow-recovery-preq}
\end{equation}
The first two cases correspond to increasing or trimming an existing position.\ The third case corresponds to opening a new long position and is therefore assigned a stricter threshold.\ Although the existing-position threshold is close to \(1/2\), note that it is not used in isolation:\ a coordinate must also pass the MAP magnitude gate and survive the ranking and implementation constraints.\ The fourth case blocks short-position initiation:\ a negative change is admissible only as a sale from an existing positive holding.\ The recovery gate is asymmetric because opening a new name is operationally different from adjusting an existing holding.\ The directional thresholds are also deliberately lower than the calm-mode threshold.\ The raw recovery set is
\begin{equation}
  S^{\mathrm{raw}}_{\mathrm{rec},h}
  =
  \left\{
  j:
  |\Delta w_{\mathrm{MAP},h,j}|\geq10^{-4},\quad
  \widehat p_{\mathrm{dir},h,j}(0)\geq p^{\mathrm{req}}_{h,j}
  \right\}.
  \label{eq:workflow-recovery-raw}
\end{equation}
The magnitude gate enforces practical size, while \(\widehat p_{\mathrm{dir},h,j}(0)\) checks sign confidence.\ The magnitude and probability filters determine which coordinates are eligible for recovery, but the implementation budget is, on purpose, limited.\ The remaining question is therefore how to prioritise eligible moves.\ We use the posterior-weighted MAP magnitude
\begin{equation}
  r_{h,j}
  =
  \widehat p_{\mathrm{dir},h,j}(0) \cdot
  |\Delta w_{\mathrm{MAP},h,j}|,
  \label{eq:workflow-recovery-rank}
\end{equation}
as a simple priority score.\ This score favours moves that are both practically
large under the MAP proposal and directionally supported by the posterior sample.\ Ranking only by \(\widehat p_{\mathrm{dir},h,j}(0)\) could allocate the recovery
budget to statistically stable but economically negligible changes, while ranking
only by \(|\Delta w_{\mathrm{MAP},h,j}|\) could favour large moves with weak
directional support.\ The product is not intended as an optimal utility rule;\ it is merely a heuristic for spending a small recovery budget on the most material posterior-supported directions.\ The approved list is then truncated to at most ten moves and at most four new buy names.

\subsection{Bounded local implementation}\label{subsec:workflow-local-implementation}

The \(\Delta w\)-MAP and posterior sample determine directions, rankings, and relative priorities; the portfolio rule determines the realised trade size.\ Let \(\Delta w^{\mathrm{impl}}_h\) denote the executed change.\ It is required to be self-financing and long-only feasible:
\begin{equation}
  \one^\top\Delta w^{\mathrm{impl}}_h=0,
  \qquad
  w^{\mathrm{old}}_h+\Delta w^{\mathrm{impl}}_h\geq0.
  \label{eq:workflow-implemented-feasibility}
\end{equation}
Negative entries of \(\Delta w^{\mathrm{impl}}_h\) are therefore permitted, but only as sales from existing positive holdings;\ they cannot create short positions.\ For any proposed update, define one-way turnover by
\begin{equation}
  \mathcal T(\Delta w)=\frac12\|\Delta w\|_1.
  \label{eq:workflow-one-way-turnover}
\end{equation}
In recovery mode, we target \(\mathcal T(\Delta w^{\mathrm{impl}}_h)=1\%\) and impose the hard cap \(\mathcal T(\Delta w^{\mathrm{impl}}_h)\leq2\%\).

Approved sells fund approved buys first.\ If the selected recovery set is buy-dominant, up to six additional weak existing holdings may be used as funders.\ The funding rule protects the current top ten holdings and any holding with weight at least 3\%, prioritises dust holdings below 0.5\%, and preserves the exact budget.\ We reiterate that at no point is the full portfolio globally renormalised.

This final implementation rule is deliberately conservative.\ It converts posterior directional evidence into a bounded local repair, leaves all other holdings unchanged, and guarantees that the realised portfolio remains long-only.\ This is the operational distinction between the proposed maintenance layer and a rolling reconstruction baseline:\ the latter is free to replace the sparse portfolio, whereas the former spends a small, pre-specified turnover budget only when realised deterioration and posterior evidence jointly support intervention.

The numerical constants in Sections~4.2--4.7 should therefore be read as an example specification for the empirical case study below, not as universal trading guidelines.\ We emphasise that the case study does not constitute a comprehensive backtesting, which, for example, would also require calibrating the length of the fitting and holding windows.\ The implementation requires explicit choices about trade execution, so the reported results are conditional on the stated trading rules, which is why we make them transparent.

 Instead, the case study aims to make one conservative proof-of-concept implementation fully auditable:\ a calibrated construction stage, posterior-informed support selection, recovery-mode directional gates, and bounded local repairs.\ In another index universe, mandate, transaction-cost environment, or desired position on the tracking-error--sparsity--turnover frontier, the same architecture could be run with different thresholds, floors, turnover budgets, recovery triggers, and trading rules.\ The contribution is the uncertainty-aware construction-and-maintenance workflow, and we do not claim that the particular numerical constants used here are generally optimal.

\section{Empirical case study}\label{sec:case-study}

\subsection{Data, protocol, metrics, and comparisons}\label{subsec:case-data}

\paragraph{Dataset and investment universe}

The empirical study uses daily returns for a fixed S\&P 500-style universe from 2020-01-03 to 2025-12-31, freely available from \emph{Yahoo Finance}.\ After calendar alignment and excluding delisted or newly listed names that are not present throughout the whole horizon, the working universe contains $p=397$ assets, and the S\&P 500 index as the benchmark.\ The fixed-universe design is deliberate:\ it isolates the statistical and decision-theoretic behaviour of the proposed sparse tracker from index-constituent entry and exit effects.\ A consequence is \emph{survivorship bias}, so in practice this may understate real-world turnover and TE when reconstitutions occur.\ We make two mitigations explicit:\ (i) we use fixed-length rolling fit windows and fixed hold periods, so that all estimates are \emph{out-of-sample} relative to the subsequent hold;\ and (ii) we report implementability proxies (turnover, active names) alongside TE.

\paragraph{Rolling-window protocol}

The rolling protocol is designed to mimic a quarterly index-tracking workflow, e.g., as in \cite{dhingra2026index}.\ The initial tracker is built from a 500-trading-day fitting window.\ It is then evaluated over the first holding window.\ Subsequent holding windows have a quarterly length.\ The resulting evaluation consists of 16 holding windows, each containing 61--65 trading days.\ All fitting windows have the same length:\ 500 trading days.

The first portfolio is constructed from the first fitting window by following Steps~1--3 of Section~4.1.\ This portfolio, henceforth referred to as \emph{the tracker}, contains 59 names after posterior support selection and simplex refitting.\ Before the first out-of-sample holding window, the tracker is adjusted once by a 0.5\% implementability floor, in the spirit of a lower holding constraint.\ This removes six additional names.\ The resulting 53-name portfolio is then held throughout the first out-of-sample holding window, H1.\ Thereafter, the portfolio is monitored at the rebalance dates and updated only when the posterior-informed rebalancing layer approves a local repair.\ Most windows do not trade.\ When deterioration appears, the method permits bounded local repairs, each targeting 1\% one-way turnover subject to a 2\% hard cap.\ The realised H1--H16 path contains five such repairs, giving total sequential turnover of 5.0\% and reducing late-horizon degradation relative to the unrepaired held tracker.

\paragraph{Rationale behind the chosen tracker}

The construction of the initial tracker is guided by the comparison protocol rather than chosen arbitrarily:\ for a fair comparison with other methods (see below), which were examined with a cardinality constraint of 45 names, we aimed for a similar target.\ Typically, for index tracking approaches that are allowed to reconstruct the tracker fully in each rebalancing exercise, the primary motivation for sparsity is to control transaction costs.\ A smaller portfolio is also desirable from a management point of view, and makes it easier to avoid illiquid assets.\ For solution methods that rely on MIP, the computational cost associated with larger portfolios is significant too.\ Such an aggressive cardinality target is less central for our proposed method, because the sequential layer does not force full reconstruction at each rebalancing date, and rarely trades.\ At the same time, allowing a substantially denser initial tracker would change the operating point of the method and could invite later deterioration through early-window overfitting.\ For these reasons, without imposing a strict cap on the number of names, we guided the construction toward 50--60 assets (\Cref{subsec:workflow-construction-grid}).

We now summarise the internal branch comparison for three initial tracker constructions, labelled by the pair
\((n_{\mathrm{target}},c^\star)\).\ The first component is the target effective support used in the initial construction-stage grid score, and the second component is the selected loss-scale multiplier.\ The three branches investigated there are therefore denoted
\((180,10)\,\), \((150,25)\), and \((130,50)\).\ All three runs use the same sequential maintenance rule:\ a one-off \(0.5\%\) implementability floor is applied before the first holding window, yielding trackers with 61, 53, and 48 names respectively;\ the conservative recovery rule then uses asymmetric posterior-directional gates for existing positions, new buys, and sells;\ approved recovery trades are ranked by posterior-weighted MAP magnitude; and implemented recovery trades target \(1\%\) one-way turnover subject to a \(2\%\) hard cap and local funding.\ Thus, the comparison isolates the effect of the initial construction branch, conditional on the same downstream maintenance rule.

To summarise these comparisons, the \((150,25)\) branch is the best-performing internal branch in this set.\ It has the lowest full-horizon mean TE, median TE, and maximum TE, despite using a mid-sized implemented support after the floor.\ The denser \((180,10)\) branch has the best early-window average, but it deteriorates more strongly in the late windows and has the largest maximum TE.\ The sparser \((130,50)\) branch trades the least, but has the weakest late-horizon tracking performance.\ This motivates using \((150,25)\) as the main case-study specification:\ it occupies the most favourable compromise among the three tested construction choices under the same conservative maintenance rule.\ The results are reported in more detail in the Supplementary Material, but some are also reported in \Cref{tab:protocol-decomposition-main}.

\paragraph{Metrics}

The construction stage uses centred returns on each fitting window, as described in \Cref{sec:workflow}.\ In contrast, all empirical tracking-error values reported in this section are computed, as is common, on realised, uncentred holding-window returns.\ This distinction is important:\ centring is a modelling device used to estimate the tracker, whereas the reported holding-window metrics correspond to the actually realised active-return path.

For holding window $h$, let $I_h$ denote the set of trading days in that window, $n_h=|I_h|$, and $w_h$ the portfolio held over the window.\ The holding-window returns are evaluated in the standard fixed-target-weight linear form \(r_t^\top w_h\);\ we do not simulate within-hold share-level weight drift or transaction-cost deductions.\ Therefore, the realised active return is
\begin{equation}
  a_t(w_h)=y_t-r_t^\top w_h,\qquad t\in I_h .
\end{equation}
The reported tracking error is the realised RMSE of active returns, expressed in basis points:
\begin{equation}
  \widehat{\TE}_h(w_h)
  =10^4\left\{\frac{1}{n_h}\sum_{t\in I_h} a_t(w_h)^2\right\}^{1/2} .
  \label{eq:hold-te}
\end{equation}
For a portfolio update from $w_{h-1}$ to $w_h$, one-way turnover is reported as
\begin{equation}
  \mathrm{TO}_h = \frac{1}{2}\|w_h-w_{h-1}\|_1,
  \label{eq:turnover}
\end{equation}
which equals total buys and total sells when the update is self-financing.\ We also distinguish weight turnover from changes in the sparse composition itself.\ Let
\[
  S_h=\{j:w_{h,j}>0\}
\]
denote the active support in holding window \(h\).\ We use the term \emph{support churn} for instability in these active sets across the rolling evaluation horizon.\ This is measured by the number of distinct names used, consecutive support overlap through the Jaccard index
\[
  J_h=\frac{|S_h\cap S_{h-1}|}{|S_h\cup S_{h-1}|},
\]
and by re-entry and short-spell diagnostics.\ Thus, turnover measures the amount traded in weight space, whereas support churn measures how often the identity of the sparse portfolio changes.\ We also report the number of active names, the number of implemented trades, and concentration summaries such as top-10 and top-25 mass.\ These quantities are essential for interpreting sparse index tracking:\ a method that reduces TE by repeatedly reconstructing the portfolio is not operationally comparable to a method that deliberately preserves composition and trades only when posterior evidence supports a bounded local repair.

The turnover measure in \eqref{eq:turnover} is used as an implementation-cost diagnostic.\ Under a proportional transaction-cost approximation, a one-way turnover of $u$ corresponds to an indicative cost drag of roughly $\kappa u$ for a cost rate $\kappa$ per dollar traded.\ This first-order approximation is standard in portfolio-optimisation treatments of turnover constraints, see, e.g.,  \cite[Sec.~6.2.5]{PalomarBook};\ but of course it does not replace a mandate-specific execution-cost model with bid--ask spreads, nonlinear market impact, fixed commissions, taxes, or liquidity constraints.

\paragraph{Comparator methods and protocol}

The external comparison set is deliberately focused.\ It is not intended to reproduce the full range of optimisation, statistical, cointegration, evolutionary, and machine-learning approaches used for sparse index tracking;\ comprehensive surveys are provided by \citet{Silva2024,dhingra2026index}.\ Instead, the comparison is organised around the decision modes most relevant to the paper.\  First, regularised-regression baselines (NNL and NNEN) represent sparse construction by non-negative shrinkage and refitting \cite{wu2014nonnegativelasso,wu2014nonnegativeEN}.\ Second, MM, MSE, and TEV represent full rolling reconstruction:\ at each
rebalance date they are allowed to rebuild the sparse portfolio from the current fitting window.\ Here MSE \cite{RuizTorrubianoSuarez2009} and TEV  \cite{DerigsNickel2003, MutungeHaugland2018} denote cardinality-constrained mean-squared tracking-error and tracking-error-variance reconstruction benchmarks, respectively, following standard sparse index-tracking objective
variants \cite{dhingra2026index}.\ MM denotes a Majorisation--Minimisation dedicated sparse-index-tracking benchmark based on empirical tracking error \cite{Benidis2018b}, computed with the \texttt{sparseIndexTracking} R package of \citet{SparseIndexTrackingR}.\ Third, additional decomposition experiments examine static holding, plug-in maintenance from external MM/MSE initial trackers, and rolling UQ reconstruction.

The cardinality target or explicit cardinality constraint used for the reconstruction comparison is $K=45$;\ for penalised regression baselines, the realised practical support may differ from exactly 45 names after numerical thresholding and refitting.\

\subsection{Final tracker performance}\label{subsec:case-performance}

\begin{figure}[h]
\centering
\includegraphics[height=5.8cm,keepaspectratio]{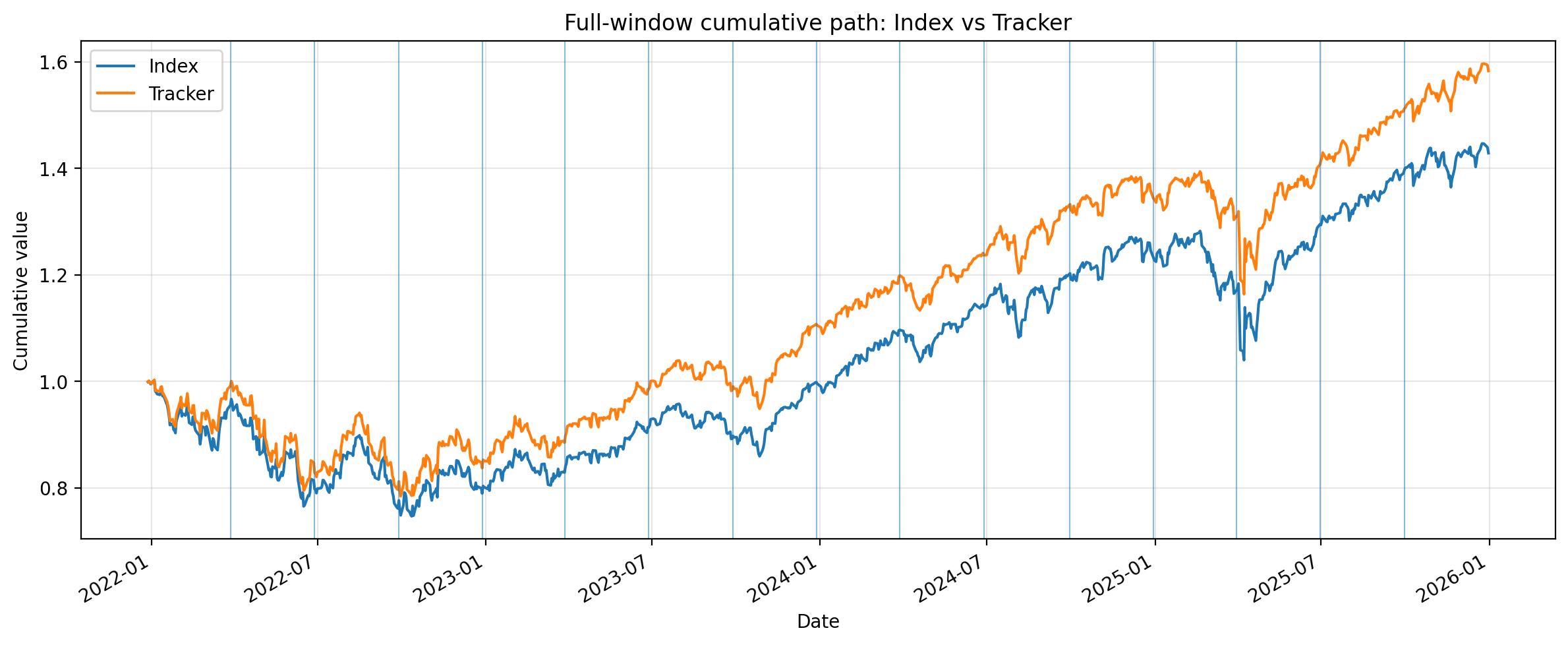}
\caption{Cumulative benchmark and tracker realised returns out-of-sample.\ The objective is tracking stability;\ cumulative outperformance is sample-specific and it is not used as a selection criterion.}
\end{figure}

\paragraph{Full-horizon summary}

\Cref{tab:final-summary} summarises the full H1--H16 evaluation.\ The mean holding-window TE is $20.556\bp$, the median is $19.409\bp$, and the maximum is $36.074\bp$.\ The early and late windows behave differently.\ Over H1--H8, the tracker records a mean TE of $15.131\bp$.\ Over H9--H16, after the market environment becomes less favourable for the initially constructed tracker, the mean TE rises to $25.981\bp$.\ This is precisely the setting in which the \emph{maintenance} layer is meant to operate:\ the initial portfolio remains sparse and stable, but local repairs are permitted once realised deterioration is detected.\ The total sequential turnover over the full horizon is $5.0\%$.\ The final portfolio contains 55 active names, with top-10 mass $45.8\%$ and top-25 mass $72.8\%$.\ Thus the final object is still a sparse, concentrated partial-replication portfolio.\ 

\begin{table}[htbp]
\centering
\caption{Final tracker full-horizon summary. Tracking-error values are realised holding-window RMSEs in basis points; turnover is total one-way sequential turnover.}
\label{tab:final-summary}
\begin{tabular}{ll}
\toprule
Metric & Value \\
\midrule
Mean holding-window TE & 20.556\bp \\
Median holding-window TE & 19.409\bp \\
Maximum holding-window TE & 36.074\bp \\
\boxit{3.06in}H1--H8 mean TE & 15.131\bp \\
H9--H16 mean TE & 25.981\bp \\
H10--H13 mean TE & 27.587\bp \\
H14--H16 mean TE & 25.680\bp \\
\boxit{3.06in}Total sequential turnover & 5.0\% \\
Final number of names & 55 \\
Final top-10 mass & 45.8\% \\
Final top-25 mass & 72.8\% \\
\bottomrule
\end{tabular}
\end{table}

\FloatBarrier

\paragraph{Hold-by-hold behaviour}

\begin{figure}[H]
\centering
\includegraphics[height=6.3cm,keepaspectratio]{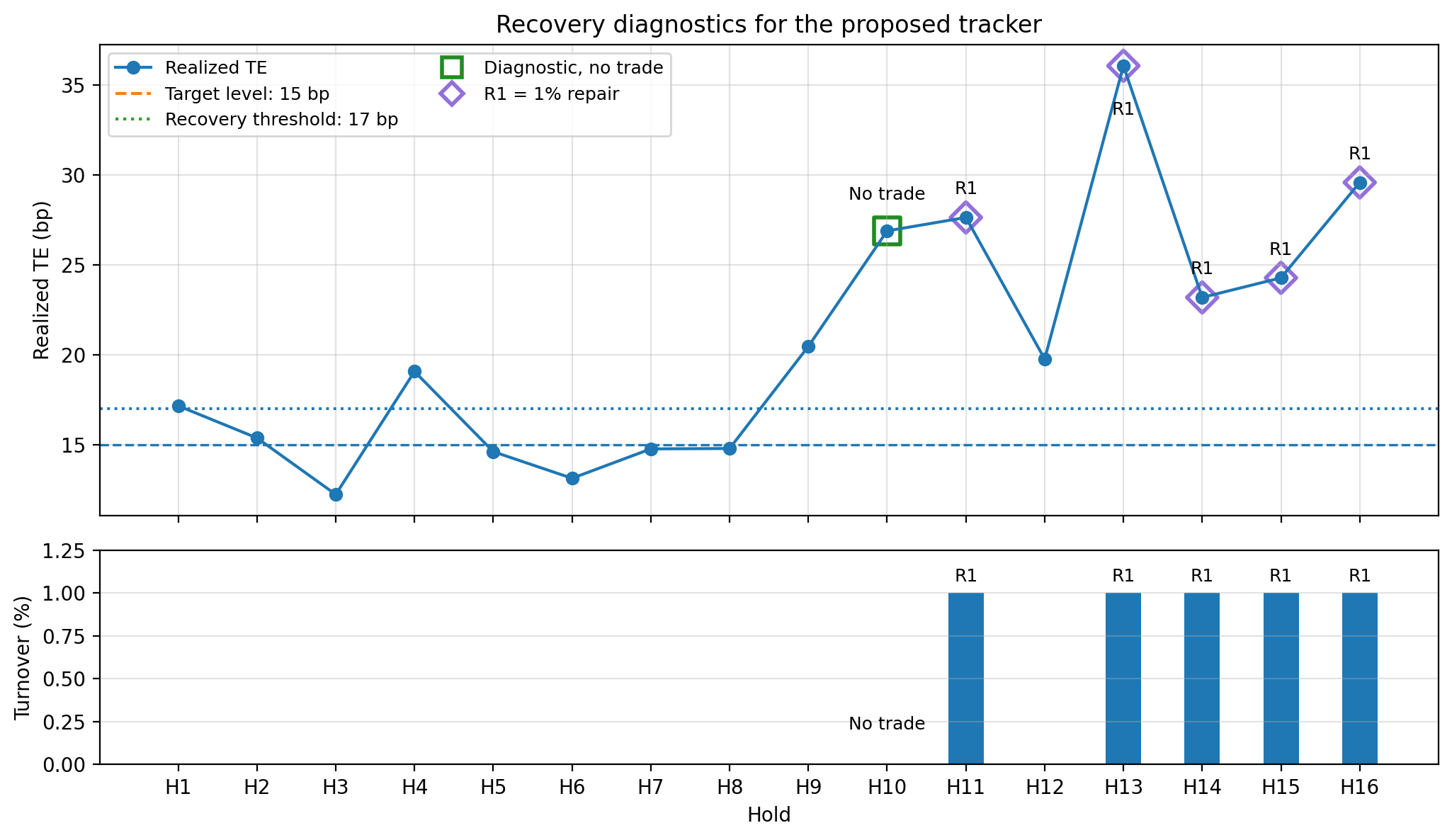}
\caption{Recovery diagnostics for the final tracker.\ The upper panel reports hold-by-hold realised tracking error for the proposed tracker, with the target and recovery-trigger bands marked.\ The lower panel reports one-way turnover implemented by the maintenance layer.}
\label{recovery-diagnostics_fig}
\end{figure}

\Cref{tab:hold-by-hold} reports the hold-level path.\ The tracker is stable and non-trading through H9.\ The first deterioration signal appears around H10, where the implementation records (e.g., \Cref{recovery-diagnostics_fig}) a sell-only recommendation but no approved destination for the released capital;\ under the stated rule no rebalance is made.\ Recovery trades then occur in H11, H13, H14, H15, and H16, each with a 1\% one-way repair.

\begin{longtable}{llllcll}
\caption{Hold-by-hold performance of the final tracker.\ Turnover is one-way and reported as a percentage.\ $n_h$ denotes the number of trading days in each window.}\label{tab:hold-by-hold}\\
\toprule
Hold & \hspace{3.5em}Dates & $n_h$ & TE (bp) & Turnover (\%) & Trades & Names \\
\midrule
\endfirsthead
\toprule
Hold & \hspace{3.5em}Dates & $n$ & TE (bp) & Turnover (\%) & Trades & Names \\
\midrule
\endhead
\bottomrule
\endlastfoot
1 & 2021-12-28--2022-03-25 & 62 & 17.151 & 0.0 & 0 & 53 \\
2 & 2022-03-28--2022-06-27 & 63 & 15.363 & 0.0 & 0 & 53 \\
3 & 2022-06-28--2022-09-27 & 64 & 12.218 & 0.0 & 0 & 53 \\
4 & 2022-09-28--2022-12-27 & 63 & 19.064 & 0.0 & 0 & 53 \\
5 & 2022-12-28--2023-03-27 & 61 & 14.599 & 0.0 & 0 & 53 \\
6 & 2023-03-28--2023-06-27 & 63 & 13.119 & 0.0 & 0 & 53 \\
7 & 2023-06-28--2023-09-27 & 64 & 14.755 & 0.0 & 0 & 53 \\
8 & 2023-09-28--2023-12-27 & 63 & 14.776 & 0.0 & 0 & 53 \\
9 & 2023-12-28--2024-03-27 & 62 & 20.460 & 0.0 & 0 & 53 \\
10 & 2024-03-28--2024-06-27 & 63 & 26.887 & 0.0 & 0 & 53 \\
11 & 2024-06-28--2024-09-27 & 64 & 27.634 & 1.0 & 7 & 52 \\
12 & 2024-09-30--2024-12-27 & 63 & 19.754 & 0.0 & 0 & 52 \\
13 & 2024-12-30--2025-03-28 & 61 & 36.074 & 1.0 & 12 & 52 \\
14 & 2025-03-31--2025-06-27 & 62 & 23.183 & 1.0 & 12 & 55 \\
15 & 2025-06-30--2025-09-29 & 64 & 24.274 & 1.0 & 12 & 55 \\
16 & 2025-09-30--2025-12-31 & 65 & 29.582 & 1.0 & 13 & 55 \\
\end{longtable}

\subsection{Recovery-mode behaviour}\label{subsec:case-recovery}
\vspace{1em}

\FloatBarrier

\paragraph{From posterior direction to implemented repair}

The recovery layer is intentionally not a direct execution of the raw $\Delta w_{\mathrm{MAP}}$ vector.\ The smoothed $\Delta w$-MAP and the MALA samples are used to identify directionally credible candidate moves.\ Candidate changes are then ranked by
\begin{equation}
  p_{\mathrm{dir},j}\,|\Delta w_{\mathrm{MAP},j}|,
\end{equation}
where $p_{\mathrm{dir},j}$ is the posterior probability of the MAP direction for coordinate $j$.\ The implemented trade is a bounded local repair:\ the tracker branch targets a 1\% one-way recovery adjustment and imposes a 2\% hard cap.\ 
Posterior computation supplies directional evidence and candidate ranking;\ the portfolio rule converts that evidence into an implementable, turnover-controlled update.

Recall from subsection \ref{subsec:workflow-recovery} that we use asymmetric recovery gates.\ Existing-position changes and sells require $p_{\mathrm{dir}}\geq0.525$, whereas new buys require $p_{\mathrm{dir}}\geq0.575$.\ The stricter new-buy threshold reflects the higher implementation burden of opening a new name.\ When approved buys dominate approved sells, weak existing holdings can be used as additional funders, subject to protection rules for the top ten holdings and positions above 3\%.\ This funding top-up is local:\ it does not renormalise the full portfolio and does not disturb large protected names.

\paragraph{Decision-level recovery events}

\Cref{tab:recovery-events} reports the decision-level events.\ The H10 diagnostic is instructive:\ the posterior evidence points to sells but provides no approved destination for the released capital, so the algorithm refuses to trade.\ This rule avoids mechanically reducing exposure without a corresponding posterior-approved buy.\ H11 and H13--H16 are buy-dominant recovery events with funding top-up.\ Each implements the 1\% one-way target;\ the number of touched names ranges from 7 to 13.

\begin{table}[htbp]
\centering
\caption{Decision-level recovery events in the full-horizon run.\ Turnover, implemented buy, and extra funding are one-way percentages.\ A dash indicates that no buy-side implementation was made.}
\label{tab:recovery-events}
\begin{tabular}{c l r r r r c c}
\toprule
Hold & Rule & Turn. & Trades & Buys & Sells & Impl. buy & Extra funding \\
\midrule
10 & SOND & 0.0 & 0 & 0 & 0 & -- & -- \\
11 & BDFT & 1.0 & 7 & 2 & 5 & 1.0 & 0.947 \\
13 & BDFT & 1.0 & 12 & 6 & 6 & 1.0 & 0.894 \\
14 & BDFT & 1.0 & 12 & 9 & 3 & 1.0 & 0.960 \\
15 & BDFT & 1.0 & 12 & 10 & 2 & 1.0 & 1.000 \\
16 & BDFT & 1.0 & 13 & 10 & 3 & 1.0 & 1.000 \\
\bottomrule
\end{tabular}
\caption*{SOND: sell-only candidates but no approved destination;\ BDFT: buy-dominant repair with local funding top-up.}
\end{table}

The H12 outcome is also informative.\ After the H11 recovery repair, the algorithm does not continue trading mechanically:\ the portfolio is held unchanged through H12, with zero additional turnover, and the realised TE falls from 27.634 bp in H11 to 19.754 bp in H12.\ This behaviour is consistent with the intended gating logic:\ recovery mode authorises bounded repairs when deterioration and posterior evidence justify intervention, but it does not force repeated trading once the held tracker has stabilised.

\FloatBarrier

\paragraph{Illustrative trades}

\Cref{tab:example-trades11} and \Cref{tab:example-trades13} show the implemented H11 and H13 trades respectively.\ The H11 repair increases NVDA and AVGO, while the necessary funding is obtained from small sales and two extra funders.\ H13 is more active and introduces AMAT as a new buy, while increasing several existing technology and infrastructure exposures (in hindsight, unsurprising given this was around the end of 2024--beginning of 2025).\ These examples show the intended character of the method:\ it identifies a small set of local moves, funds them without a global rebalance, and leaves the rest of the portfolio untouched.

\begin{table}[htbp]
\centering
\caption{Illustrative implemented trades in H11.\ Weights and changes are in percent.}
\label{tab:example-trades11}
\begin{tabular}{cllccr}
\toprule
Hold & Ticker & Status/action & Current weight & Implemented change & $p_{\mathrm{dir}}$ \\
\midrule
11 & NVDA & existing/buy & 1.256 & 0.681 & 0.568 \\
11 & AVGO & existing/buy & 2.425 & 0.319 & 0.540 \\
11 & CMCSA & existing/sell & 1.964 & -0.013 & 0.528 \\
11 & GOOG & existing/sell & 6.519 & -0.019 & 0.530 \\
11 & GL & existing/sell & 1.290 & -0.021 & 0.550 \\
11 & HOLX & extra funder & 0.777 & -0.434 & 0.506 \\
11 & INCY & extra funder & 0.513 & -0.513 & 0.512 \\
\bottomrule
\end{tabular}
\caption*{{\it Note:} For extra funders, $p_{dir}$ is reported diagnostically;\ these names are selected by the local funding rule and are not required to pass the posterior sell gate.\ In the ``Status/action'' column, ``existing'' means that the name is already held before the recovery trade, while ``buy'' and ``sell'' denote the implemented direction of the change.}
\end{table}

\begin{table}[htbp]
\centering
\caption{Illustrative implemented trades in H13.\ Weights and changes are in percent.}
\label{tab:example-trades13}
\begin{tabular}{cllccr}
\toprule
Hold & Ticker & Status/action & Current weight & Implemented change & $p_{\mathrm{dir}}$ \\
\midrule
13 & NVDA & existing/buy & 1.937 & 0.332 & 0.642 \\
13 & AVGO & existing/buy & 2.744 & 0.183 & 0.569 \\
13 & AMAT & new/buy & 0.000 & 0.177 & 0.585 \\
13 & AMD & existing/buy & 0.945 & 0.123 & 0.548 \\
13 & PWR & existing/buy & 1.209 & 0.113 & 0.545 \\
13 & HPE & existing/buy & 1.212 & 0.071 & 0.542 \\
13 & ARE & existing/sell & 1.356 & -0.020 & 0.534 \\
13 & JNJ & existing/sell & 4.138 & -0.023 & 0.540 \\
13 & AMT & existing/sell & 1.779 & -0.027 & 0.542 \\
13 & GL & existing/sell & 1.269 & -0.037 & 0.554 \\
13 & HOLX & extra funder & 0.343 & -0.343 & 0.523 \\
13 & ORLY & extra funder & 0.856 & -0.551 & 0.503 \\
\bottomrule
\end{tabular}
\end{table}

\paragraph{Interpretation}

The recovery layer is deliberately conservative.\ Given the existing tracker, it activates only for sufficient posterior directional evidence that justifies a small self-financing repair.\ The empirical behaviour is consistent with that design.\ Most windows do not trade;\ when deterioration appears, the method deploys bounded 1\% repairs that reduce the late-horizon degradation relative to the unrepaired held tracker, while keeping total sequential turnover at $5.0\%$.

\paragraph{Sensitivity to more active recovery rules}
More active recovery policies are examined in the Supplementary Material.\ These variants use the same posterior-informed maintenance engine, but allow combinations of larger repair budgets, more approved moves, sell-only destination rules, and, in one case, an additional fit-window escalation trigger.\ They improve late-horizon tracking error in this sample, but at the expected cost of higher turnover, greater rule complexity, and more concentrated final portfolios.\

\FloatBarrier

\subsection{Comparisons with other methods}
\label{sec:case-competitors}

\vspace{1em}

Before comparing against rolling reconstruction baselines, it is useful to separate four evaluation modes.\ \Cref{tab:protocol-decomposition-main} summarises the additional experiments.\ We first explain the four modes we compare below. 

\begin{itemize}

\item Rows 1 and 2 of \Cref{tab:protocol-decomposition-main} record the performance of the ``main'' (150,25) tracker (resulting in 53 names) maintained with our proposed rebalancing pipeline, as well as the sparser (130,50) tracker (resulting in 48 names) designed and maintained with the same approach.\ More comparisons with the second tracker can be found in the Supplementary Material.\

\item Static holding (row 3) shows the cost of doing nothing after the first fit window, i.e., buying and holding the (150,25) tracker (designed with our proposed UQ-gating mechanism) but without maintenance/rebalancing after the first hold.\ 

\item We used the MSE and MM methods to construct two 45-asset tracking portfolios with the same data of the original fitting window but with no other intervention.\ We then plugged each of the two trackers into our maintenance pipeline under exactly the same rules as our proposed tracker.\ The plug-in maintenance exercise shows (rows 4 and 5) that the $\Delta w$-layer is portable to external starting trackers, although the gains are modest.\ 

\item Finally, we implemented a rolling reconstruction for our tracker (row 6), by which we mean that before each new holding period we construct a new portfolio based on the hybrid (optimisation+UQ gating) approach we took for the first fit-window.\ This rolling reconstruction shows that the UQ-gated construction layer itself is competitive in the low-TE reconstruction regimes, but only when it accepts reconstruction-level turnover and support churn, as observed in the corresponding experiments with the other methods (see  \Cref{tab:competitors} and \Cref{tab:support_churn_competitors_vertical}).\
\end{itemize}

\begin{table}[htbp]
\centering
\caption{Protocol-decomposition experiments.\ TE values are realised holding-window RMSEs in basis points;\ turnover is total sequential one-way turnover.}
\label{tab:protocol-decomposition-main}
\small
\begin{tabular}{lrrrrl}
\toprule
Path & Mean TE & Median TE & Max TE & Turnover\\
\midrule
(150,25)-Tracker & 20.556 & 19.409 & 36.074 & 5.0\%\\
(130,50)-Tracker & 23.942 & 20.011 & 39.994 & 2.0\%\\
(150,25)-Static hold & 22.704 & 19.762 & 42.192 & 0.0\% \\
MSE-with-maintenance & 21.957 & 20.062 & 36.742 & 4.0\%\\
MM-with-maintenance & 24.637 & 23.139 & 44.840 & 4.0\%\\
Tracker rolling reconstruction & 17.711 & 16.825 & 23.153 & 604.6\% \\
\bottomrule
\end{tabular}
\end{table}

\FloatBarrier

This comparison in particular suggests that our main result (low-turnover) comes from the state-aware maintenance formulation, not from a weak reconstruction benchmark.

\paragraph{Full-reconstruction baselines}

\Cref{tab:competitors} compares the final tracker with the full ``competitor'' set.\ MM, MSE, and TEV form the strongest tracking-error group.\ Their mean TE values are around $17.3\bp$, compared with $20.556\bp$ for the proposed tracker.\ 

\begin{table}[htbp]
\centering
\caption{Full-horizon comparison with competitors.\ TE values are realised holding-window RMSEs in basis points.\ Turnover is total one-way turnover over all holding windows.}
\label{tab:competitors}
\begin{tabular}{lrrrrrr}
\toprule
Metric 
& Tracker 
& MM
& MSE 
& NNEN 
& NNL 
& TEV \\
\midrule
Mean TE        & 20.556 & 17.261 & 17.265 & 25.148 & 31.321 & 17.374 \\
Median TE      & 19.409 & 17.015 & 17.583 & 24.270 & 30.253 & 16.692 \\
Max TE         & 36.074 & 23.900 & 23.470 & 42.077 & 50.456 & 22.468 \\
Total turnover & 5.0\%  & 492.3\% & 663.4\% & 403.1\% & 443.2\% & 680.0\% \\
Mean active names     & 53.2     & 44.2    & 45.0    & 34.9    & 34.5    & 45.0 \\
\bottomrule
\end{tabular}
\end{table}

\FloatBarrier

The performance gap opens mainly after Holds 8--9.\ Up to that point, the TE performance of the proposed tracker is on par with the strongest competitor group (see \Cref{tab:competitor_te_h1_h9_summary}), and this is achieved without a single rebalancing trade, in stark contrast with the other methods.

\begin{table}[htbp]
\centering
\caption{Summary of realised tracking error over Holds 1--9. TE values are holding-window RMSEs in basis points.}
\label{tab:competitor_te_h1_h9_summary}
\begin{tabular}{lrrrrrr}
\toprule
Metric & Tracker & MM & MSE & NNEN & NNL & TEV \\
\midrule
Mean TE   & 15.723 & 15.250 & 15.684 & 25.255 & 30.729 & 16.154 \\
Median TE & 14.776 & 14.590 & 14.050 & 25.849 & 29.619 & 15.436 \\
Max TE    & 20.460 & 19.054 & 20.257 & 42.077 & 43.366 & 20.410 \\
\bottomrule
\end{tabular}
\end{table}

\begin{figure}[H]
\centering
\includegraphics[height=5.8cm,keepaspectratio]{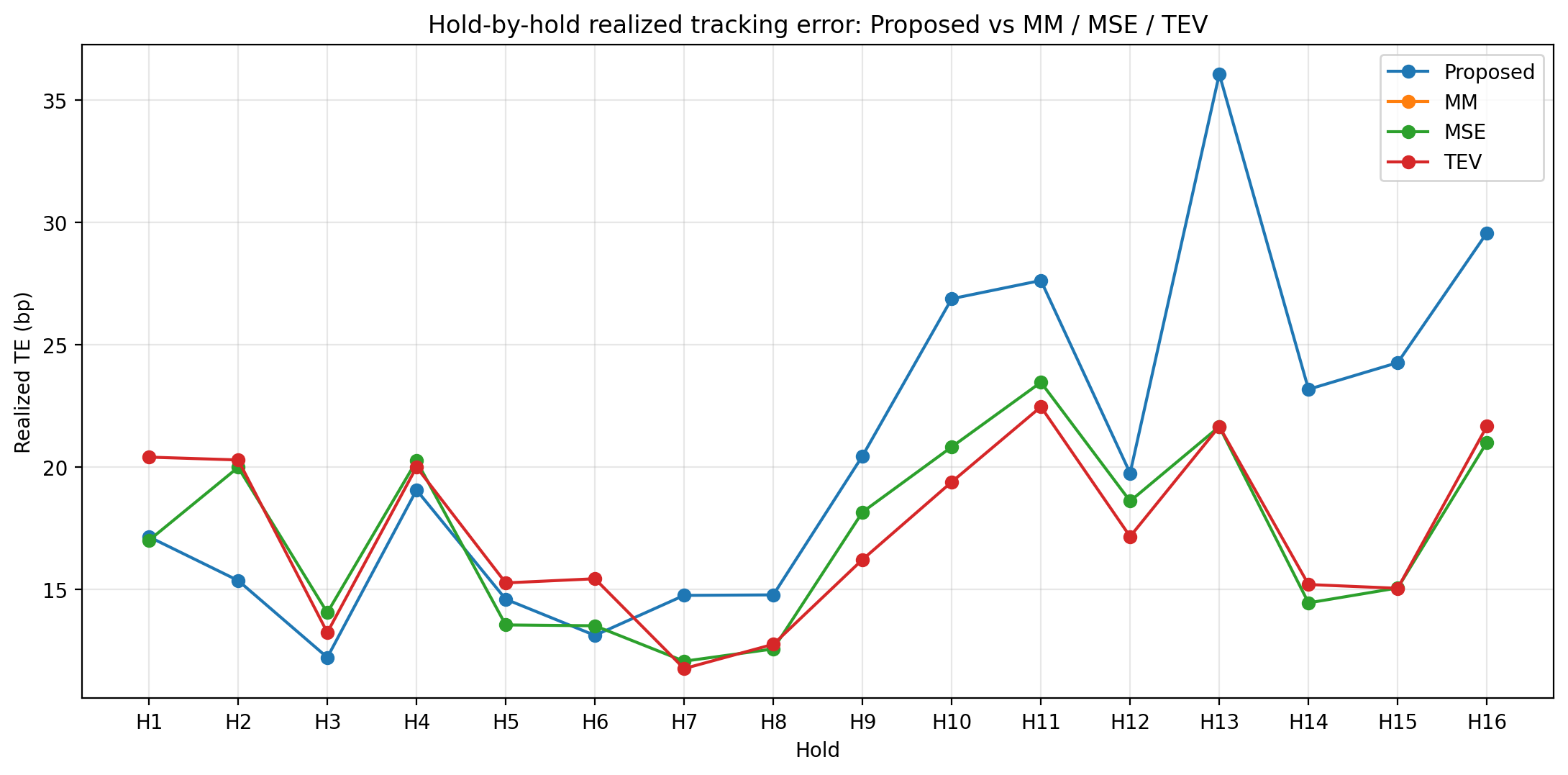}
\caption{Hold-by-hold TE comparison vs main reconstruction baselines.}
\end{figure}

The turnover contrast is substantial.\ The proposed tracker ends up using $5.0\%$ total sequential turnover across all 16 holding windows;\ MM uses $492.3\%$, MSE uses $663.4\%$, and TEV uses $680.0\%$.\ The proposed method therefore occupies the low-turnover extreme of the empirical frontier.\ It gives up (with this rule set) approximately three basis points of mean holding-window TE relative to the best reconstruction baselines, but avoids repeated wholesale portfolio reconstruction.

A further distinction is revealed by the support-churn diagnostics defined in Section~5.1.\ The reconstruction baselines exhibit substantial instability in their active sets.\ For example, the MSE and TEV $(K=45)$ baselines hold exactly 45 assets per window, but touch 225 and 228 distinct assets, respectively, over the 16-window horizon.\ Their mean consecutive support Jaccard indices are only 0.353 and 0.339.\ Moreover, many support changes are not permanent substitutions:\ MSE produces 147 re-entry events, and TEV produces 154, where a re-entry is an asset that is dropped and subsequently bought back.\ In both cases, the median absence before re-entry is two holding windows.\ Thus, the high turnover of the reconstruction baselines reflects not only large trading volume but also repeated support reversal.\ This is precisely the form of support churn that our proposed uncertainty-gated repair mechanism is designed to avoid.

\begin{table}[htbp]
\centering
\caption{Support-churn diagnostics for the \(K=45\) reconstruction and sparse-regression competitors.\ TE values are realised holding-window RMSEs in basis points. Turnover is total one-way turnover over the full 16-window horizon.}
\label{tab:support_churn_competitors_vertical}
\scriptsize
\begin{tabular}{lrrrrr}
\toprule
Metric 
& MSE
& MM
& TEV
& NNEN
& NNL \\
\midrule
Mean TE          & 17.265  & 17.261  & 17.374  & 25.148  & 31.321 \\
Turnover         & 663.4\% & 492.3\% & 680.0\% & 403.1\% & 443.2\% \\
Mean names       & 45.0    & 44.2    & 45.0    & 34.9    & 34.5 \\
Unique names     & 225     & 190     & 228     & 107     & 103 \\
Mean \(J_k\)     & 0.353   & 0.505   & 0.339   & 0.630   & 0.604 \\
Re-entries       & 147     & 77      & 154     & 49      & 60 \\
Fast re-entries  & 97/147  & 51/77   & 95/154  & 32/49   & 44/60 \\
Short spells     & 243/372 & 145/267 & 248/382 & 60/156  & 61/163 \\
\bottomrule
\end{tabular}
\begin{flushleft}
\footnotesize
Notes:\ The support \(S_k\) and consecutive support Jaccard index \(J_k\) are defined in Section~5.1.\
A re-entry is an asset that is absent from \(S_{k-1}\), present in \(S_k\), 
and had appeared in some earlier support 
\(S_\ell\), \(\ell<k-1\).\ 
``Fast re-entries'' are re-entries after at most three absent holding windows.\ 
``Short spells'' reports single-window holding spells divided by total support spells.
\end{flushleft}
\end{table}


\section{Discussion}

The empirical study demonstrates an uncertainty-aware approach to sparse tracker maintenance, rather than a comprehensive backtest of an investable trading strategy.\ The construction stage produces an implementable sparse tracker by combining an effective generalised-Bayes loss scale, empirical-Bayes shrinkage calibration, posterior practical-positivity screening, and long-only refitting.\ The sequential layer then changes the decision variable from the full portfolio to the self-financing trade $\Delta w$.\ Consequently, the default action is to preserve the existing tracker, and intervention is allowed only when realised tracking deterioration and posterior directional evidence jointly support a local repair.

The main empirical message is a tracking-error--turnover--persistence trade-off.\ The strongest reconstruction baselines, especially MM, MSE, and TEV, achieve lower average and maximum realised tracking error in this case study.\ They do so by rebuilding the sparse portfolio at every rebalance date, generating several hundred per cent total turnover and substantial support churn.\ By contrast, the proposed tracker gives up tracking precision relative to that high-turnover reconstruction regime, but uses only $5\%$ total sequential one-way turnover and preserves a stable sparse composition.\ Under proportional cost approximations, turnover is a transparent first-order proxy for implementation cost;\ hence the TE reduction achieved by reconstruction should be interpreted jointly with the trading intensity required to obtain it.

The additional experiments clarify what should, and should not, be inferred from the comparison.\ ``Static hold'' controls show that the initial tracker can remain competitive when left untouched, but also that the proposed maintenance layer reduces late-horizon deterioration.\ Plug-in experiments starting from MM and MSE portfolios show that the $\Delta w$ layer is not tied to the UQ construction stage:\ it can be attached to externally supplied sparse trackers and produces improvements over static holding.\ The best low-turnover path is nevertheless obtained when our UQ-constructed tracker is paired with the UQ maintenance layer, suggesting that the initial state being maintained matters.\ A plausible explanation is that posterior support screening removes some fragile in-sample names before the maintenance process begins;\ at this stage, this comment should be treated as an interpretation of the present evidence, not as a proved mechanism.

The rolling UQ reconstruction experiment is also important.\ When the UQ construction procedure is re-run independently on every fitting window, it achieves mean/median/maximum TE of $17.711/16.825/23.153$ bp, close to the strongest MM/MSE/TEV reconstruction baselines.\ However, it uses $604.6\%$ total one-way turnover and touches 213 distinct names.\ This does not contradict the robustness interpretation of the initial UQ support gate.\ It shows that posterior support screening alone does not create multi-period state persistence.\ If the construction problem is solved from scratch at each date, even an uncertainty-aware construction method behaves like a reconstruction method.\ The low-turnover advantage comes from the $\Delta w$ maintenance formulation, which explicitly treats the currently-held tracker as the state.

This distinction is important for positioning the work relative to rebalancing literature.\  Existing dynamic and multi-period index-tracking papers commonly incorporate transaction costs, cash flows, lot constraints, or rebalancing triggers, and often evaluate internal variants of the proposed optimisation model.\ The present contribution is narrower but complementary:\ it provides an auditable posterior-gated maintenance layer for a sparse tracker.\ Candidate repairs must have a non-negligible MAP direction, posterior directional support, feasible local funding, long-only feasibility, and an explicit turnover budget.\ These requirements make the rebalancing decision inspectable and suppress many trades that would be natural under full reconstruction.

\section{Conclusion}

This paper proposes a generalised-Bayes computational workflow for sparse index tracking and portfolio maintenance.\ The method separates decisions that are often conflated:\ calibrating an effective tracking-loss scale, estimating a continuous sparse model, converting posterior information into an implementable support, and deciding whether an existing tracker should be locally repaired.\ The resulting pipeline combines proximal optimisation, empirical-Bayes SAPG calibration, proximal MCMC uncertainty summaries, and posterior-gated trading rules.

The proposed method is not intended to be a universal replacement for full reconstruction baselines.\ The latter are strong tracking optimisers, but their standard rolling deployment repeatedly replaces the sparse portfolio.\ It is precisely this portfolio-management mode that the proposed method is designed to address:\ our distinctive methodological contribution is in the incumbent-state $\Delta w$ maintenance policy, where the currently-held sparse tracker is treated as a state variable.\ Posterior information is then used to decide whether the state should be preserved or locally repaired.\ This makes uncertainty visible at the two points where it is most operationally useful:\ selecting the initial effective support and deciding whether proposed trades are sufficiently directionally supported to justify spending turnover.
 
Our method improves on static holding of the same initial tracker, improves substantially on non-negative regularised-regression baselines, and uses only a small fraction of the turnover generated by rolling reconstruction methods.\ The plug-in experiments indicate that the maintenance layer is portable to externally constructed sparse trackers, while the rolling UQ reconstruction experiment shows that the UQ construction machinery is itself competitive as a reconstruction engine, but high-turnover when used that way.

Future work should test the framework on explicit transaction-cost models, and mandate-specific constraints such as sector, liquidity, and upper/lower-holding rules.\ It would also be useful to study adaptive recovery policies more systematically (see the Supplementary Material for encouraging results), including more active recovery triggers and links to turnover-sparsity formulations.\ Another interesting and practically relevant extension would combine scheduled reviews with event-driven inspections based on active-return accumulation, TE deterioration, volatility regimes, or posterior predictive probabilities of breaching a TE band, leading to a joint timing-and-sizing formulation for sparse-tracker maintenance.\ Last, the realised cumulative-return paths in the above case study also suggest a possible connection with enhanced index tracking;\ although this was not an optimisation objective and should be treated as sample-specific.\ Therefore, exploring the adaptation of this pipeline for enhanced index tracking is a natural direction for future work.

\section*{Acknowledgements:} The author wishes to thank K.\ Triantafyllopoulos (University of Sheffield) and K. Zygalakis (University of Edinburgh) for useful discussions. 

\bibliographystyle{apacite}
\bibliography{references}

\appendix

\renewcommand{\thesubsection}{\Alph{section}.\arabic{subsection}}
\renewcommand{\thesubsubsection}{\Alph{section}.\arabic{subsection}.\arabic{subsubsection}}

\section{Computational summary: proximal MCMC, Moreau--Yosida smoothing, posterior sampling, and SAPG calibration}
\label{app:computational-summary}

This appendix summarises the computational layer used to calibrate the global sparsity rate and to compute the posterior summaries used by the support-selection and rebalancing rules.\ The full diagnostic record, including grid tables, trace plots, effective sample sizes, Monte Carlo standard errors, support-sensitivity tables, and rebalancing-decision audits, is reported in the Supplementary Material.

\subsection{Proximal MCMC and MAP computation}
\label{app:proximal-summary}

Both the construction and rebalancing stages lead to composite objectives with a smooth quadratic term and a nonsmooth weighted-\(\ell_1\) term.\ In the construction stage, for a fixed effective loss scale \(\omega=1/\sigma_{\mathrm{eff}}^2\) and sparsity rate \(\theta\), the nonsmooth component is
\[
G_\theta(w)
=
\theta \sum_{j=1}^p \alpha_j |w_j|.
\]
The corresponding Moreau--Yosida envelope with smoothing parameter \mbox{\(\lambda_{\mathrm{MY}}>0\)} is
\[
G_{\theta,\lambda_{\mathrm{MY}}}(w)
=
\min_{z\in\mathbb{R}^p}
\left\{
G_\theta(z)
+
\frac{1}{2\lambda_{\mathrm{MY}}}\|z-w\|_2^2
\right\},
\]
with gradient
\[
\nabla G_{\theta,\lambda_{\mathrm{MY}}}(w)
=
\frac{1}{\lambda_{\mathrm{MY}}}
\left\{
w-\operatorname{prox}_{\lambda_{\mathrm{MY}}G_\theta}(w)
\right\}.
\]
For the construction-stage weighted-\(\ell_1\) penalty, the proximal map is component-wise weighted soft-thresholding:
\[
\left[
\operatorname{prox}_{\lambda_{\mathrm{MY}}G_\theta}(w)
\right]_j
=
\operatorname{sign}(w_j)
\max\left\{
|w_j|-\lambda_{\mathrm{MY}}\theta\alpha_j,0
\right\}.
\]

The rebalancing stage uses the same proximal idea, but the state variable is the self-financing change \(u=\Delta w\).\ The nonsmooth term is
\[
G_{\kappa}^{\Delta}(u)
=
\kappa \sum_{j=1}^p \alpha^{(\Delta)}_j |u_j|
+
\iota_{\{1^\top u=0\}}(u),
\]
where \(\iota_C\) denotes the indicator of the set \(C\).\ Its proximal map is
\[
\operatorname{prox}_{\lambda_{\mathrm{MY}}G_\kappa^\Delta}(z)
=
\arg\min_{u:1^\top u=0}
\left\{
\frac12\|u-z\|_2^2
+
\lambda_{\mathrm{MY}}\kappa
\sum_{j=1}^p \alpha_j^{(\Delta)}|u_j|
\right\}.
\]
In our implementation, this is computed by weighted soft-thresholding with a scalar shift chosen so that the output sums to zero.\ This is the main computational difference between construction-stage posterior sampling and \(\Delta w\)-posterior sampling.

The MAP estimates reported in the main workflow are computed by proximal-gradient/FISTA-type iterations applied to the corresponding smoothed objective.\ The final investable portfolio is not the raw smoothed MAP:\ the MAP and posterior sample are used to identify support or trade directions, after which the portfolio is refitted or locally implemented under the long-only and budget constraints described in the main text.

\subsection{Posterior sampling by preconditioned MALA}
\label{app:mala-summary}

Let \(\Phi_{\lambda_{\mathrm{MY}}}\) denote the smoothed negative log posterior, including the smooth tracking term, any soft-budget term, and the Moreau--Yosida-smoothed nonsmooth component.\ Posterior summaries in the main branch are computed using a preconditioned Metropolis-adjusted Langevin proposal of the form
\[x'=x- \gamma P^2\nabla \Phi_{\lambda_{\mathrm{MY}}}(x)+ \sqrt{2\gamma}\,P\xi,
\qquad
\xi\sim\mathcal{N}(0,I),
\]
followed by the standard Metropolis--Hastings accept/reject step for the smoothed target.\ In the construction stage, \(x=w\);\ in the maintenance stage, \(x=u=\Delta w\).

The diagonal preconditioner is chosen as a Jacobi approximation to the local quadratic curvature.\ For the construction sampler, this includes the data-fit curvature and the soft-budget contribution.\ For the \(\Delta w\) sampler, the self-financing constraint is handled through the proximal map, so the diagonal preconditioner is based on the data-Hessian part on the current fitting window.\ The MALA step size is tuned during a short adaptation phase and then fixed for the reported long run.\ The Supplementary Material reports the retained sample sizes, acceptance rates, ESS values, MCSEs, and representative autocorrelation diagnostics used to check that the posterior probabilities entering the decision rules are numerically stable and reflect an effective exploration of the posterior landscape.\

\subsection{SAPG calibration of the sparsity rate}
\label{app:sapg-summary}

For a fixed effective loss scale, the construction-stage weighted-Laplace rate \(\theta\) is calibrated by stochastic approximation proximal gradient (SAPG).\ Let
\[
g(w)=\sum_{j=1}^p \alpha_j |w_j|.
\]
For the weighted-Laplace family, the corresponding log-scale score has the form
\[
p-\theta\,\mathbb{E}_{\pi_{\omega,\theta}}\{g(W)\},
\]
where the expectation is taken under the current generalised-Bayes posterior.\ SAPG replaces this expectation by a Markov chain sample generated from the current smoothed posterior.\ Writing \(\eta=\log\theta\), the update is
\[
\eta_{k+1}
=
\Pi_{[\log\theta_{\min},\log\theta_{\max}]}
\left[
\eta_k
+
\rho_k
\left\{
p-\theta_k g(w^{(k)})
\right\}
\right],
\qquad
\theta_k=\exp(\eta_k),
\]
where \(w^{(k)}\) is the current MYULA state and \(\rho_k\) is a Robbins--Monro step size.\ After burn-in, the empirical-Bayes estimate is computed by Polyak--Ruppert averaging on the log scale:
\[
\bar\eta
=
\frac{\sum_{k>k_{\mathrm{burn}}}\omega_k\eta_k}
{\sum_{k>k_{\mathrm{burn}}}\omega_k},
\qquad
\theta_{\mathrm{EB}}=\exp(\bar\eta).
\]

The \(\Delta w\) rebalancing posterior uses the same principle, but the sparsity rate is denoted by \(\kappa\) and the state is restricted to the self-financing hyperplane.\ If
\[
S_\Delta(u)
=
\sum_{j=1}^p \alpha_j^{(\Delta)} |u_j|,
\]
then the corresponding log-scale update uses the dimension \(p-1\):
\[
\zeta_{k+1}
=
\Pi_{[\log\kappa_{\min},\log\kappa_{\max}]}
\left[
\zeta_k
+
\rho_k
\left\{
(p-1)-\kappa_k S_\Delta(u^{(k)})
\right\}
\right],
\qquad
\kappa_k=\exp(\zeta_k).
\]
The change from \(p\) to \(p-1\) reflects the self-financing constraint \(1^\top u=0\).

The SAPG calibration is conditional on the effective loss scale selected by the construction or rebalancing grid.\ Thus, the workflow should not be interpreted as estimating a physical residual variance.\ The grid selects an effective generalised-Bayes learning rate, and SAPG calibrates the sparsity rate conditional on that learning rate.\ The Supplement gives the selected construction and rebalancing grid rows, the corresponding effective MAP penalty scales, and the posterior diagnostics for the chains used in the support and trade decisions.
\end{document}